 \documentclass[preprint,aps]{revtex4}
\usepackage[pdftex]{graphicx}
 
\begin{document}
  \newcommand{\Qed}{\rule{2.5mm}{3mm}}
 \newcommand{\balpha}{\mbox{\boldmath {$\alpha$}}}
 \def\Tr{{\rm Tr}}
 \def\(#1)#2{{\stackrel{#2}{(#1)}}}
 \def\[#1]#2{{\stackrel{#2}{[#1]}}}
 \def\A{{\cal A}}
 \def\B{{\cal B}}
 \def\Sb#1{_{\lower 1.5pt \hbox{$\scriptstyle#1$}}}
 \draft 
%\documentstyle[preprint,aps]{revtex}
%\begin{document}
%\draft
\title{Does dark matter consist of baryons of new stable family quarks?\\%
}

\author{G. Bregar, N.S. Manko\v c Bor\v stnik}
\address{Department of Physics, FMF, University of
Ljubljana, Jadranska 19, 1000 Ljubljana}

\date{\today}

%\maketitle

\begin{abstract} 
%$\[-]{56} = $\\
%$\Tr(\[\pm] {56} ) = 1 $
We investigate the possibility that the dark matter consists of clusters of the   
heavy family quarks and leptons with zero Yukawa couplings to the lower families. 
Such a family is predicted by the {\it approach unifying spin and charges} as the 
fifth family.   
We make a rough estimation of properties of  baryons  of this new family members, 
of their behaviour during the evolution of the universe and when scattering on 
the ordinary matter and study possible limitations on the family properties due 
to the cosmological and direct experimental evidences. 
\end{abstract}

\maketitle
%\end{document}

%
\section{Introduction}
\label{introduction}

Although the origin of the dark matter is 
unknown, its gravitational interaction with the known matter  and other cosmological 
observations 
require from the candidate for the dark matter constituent that: 
i. The scattering amplitude of a cluster of constituents with the ordinary matter 
and among the dark matter clusters themselves must be small enough, % to be in agreement 
% with the observations 
so that no effect of such scattering has been observed, except possibly 
in the DAMA/NaI~\cite{rita0708} and not (yet?) in the CDMS and other experiments~\cite{cdms}. 
ii. Its  density distribution (obviously different from the ordinary matter density 
distribution) causes that all the stars  within a galaxy rotate approximately with the same
velocity (suggesting that the density is approximately spherically symmetrically distributed, 
descending with the second power of the distance from the center, it is extended also far 
out of the galaxy,  manifesting the gravitational lensing by galaxy clusters).  
iii. The dark matter constituents must be stable in comparison with the age of our universe,  
having obviously for many orders of magnitude different time scale for forming (if at all)  
solid matter than the ordinary matter. 
iv. The dark matter constituents had to be formed during the evolution of our 
universe %(with the inflationary development included) 
so that they contribute today the main part of the matter ((5-7) times as much as the 
ordinary matter).

There are several candidates for the massive dark matter constituents in the literature, like, for example,
WIMPs (weakly interacting massive particles), the references can be found 
in~\cite{dodelson,rita0708}. 
In this paper we discuss the possibility that the dark matter constituents are 
clusters of a stable (from the point of view of the age of the universe)
family of quarks and leptons. Such a family is predicted 
by the approach unifying spin and  charges~\cite{pn06,n92,gmdn07}, proposed by one 
of the authors of this paper: N.S.M.B. This approach is showing a new way beyond the standard model of the 
electroweak  and colour interactions by answering the open questions of this model like: Where do 
the families originate?, Why do only the left handed quarks and leptons carry the weak charge, 
while the right handed ones 
do not? Why do particles carry the observed $SU(2), U(1)$ and $SU(3)$ charges? Where does the Higgs field 
originate from?, and others. 

 There are several 
attempts in the literature trying to understand the origin of families. All of them, however, 
in one or another way (for example through choices of appropriate groups) simply postulate that there are at 
least three families, as 
does the standard model of the electroweak and colour interactions. 
Proposing the (right) mechanism for generating families is to our understanding  
the most promising guide to   
physics beyond the standard model. 

{\it The approach unifying spin and charges is 
offering the mechanism for the appearance of families.} It introduces the {\it second 
kind}~\cite{pn06,n92,n93,hn02hn03} 
of the Clifford algebra objects,  which generates families as the  
{\it equivalent representations 
to the Dirac spinor representation}. The references~\cite{n93,hn02hn03} show that there are 
two, only two, kinds of the Clifford algebra objects, one used by Dirac to describe the spin of fermions. 
The second kind forms the equivalent representations with respect to the Lorentz group for spinors~\cite{pn06} and  
the families do form the equivalent representations with respect to the Lorentz group. 
The approach, in which fermions carry two kinds of spins (no charges), predicts from the simple starting action 
 more than the observed three families. It predicts 
two times  four families with masses several orders of magnitude bellow the 
unification scale of the three observed charges. 

Since due to the approach 
(after assuming a particular, but to our opinion trustable, way of a nonperturbative breaking of the starting symmetry) 
the fifth family decouples in the Yukawa couplings from 
the lower four families (whose the fourth family quark's  mass is predicted to be at around $250$ GeV or 
above~\cite{pn06,gmdn07}),    
the fifth family quarks and leptons are stable as required by the condition iii.. 
Since the masses of all the members of the fifth family lie, due to the approach, much above the known three  
and the predicted fourth family masses,  
%(the fourth family might according to the first very rough estimates of the approach be even seen at the LHC), 
the baryons made out of the fifth family form small enough  clusters (as we shall see in section~\ref{properties}) so that 
 their scattering amplitude among themselves and with the ordinary matter  
is small enough and also the number of clusters (as we shall see in  section~\ref{evolution})  is low enough to fulfil 
the conditions i. and iii.. 
Our study of the behaviour of the fifth family quarks in the cosmological evolution 
(section~\ref{evolution}) shows that also the condition 
iv. is fulfilled, if the fifth family masses are large enough.
%%%%%******

Let us add that there are several assessments about masses of a possible (non stable) fourth family of quarks and leptons, 
which follow from the analyses of the existing experimental data and the cosmological observations. 
Although most of physicists have  doubts about the existence of any more 
than the three observed families, the analyses clearly show that neither the experimental electroweak data~\cite{okun,pdg}, 
nor the cosmological observations~\cite{pdg} forbid the existence of more than three 
%%%%
%%%%KOMENTAR ZELI REFEREE
%%%%
families, as long as the masses of the fourth family quarks are higher than a few hundred 
GeV and the masses of the fourth family leptons above one hundred GeV ($\nu_4$ could be above $50$ GeV). 
 We studied in the ¨references~\cite{pn06,gmdn07,n09} possible (non perturbative)  breaks of the  symmetries
 of the simple starting Lagrangean which, by predicting the Yukawa couplings, leads at low energies first 
to twice four families with no Yukawa couplings between these two groups of families. One group 
obtains at the last break masses of several hundred TeV or higher, 
while the lower four families stay massless and mass protected~\cite{n09}.
For one choice of the next break~\cite{gmdn07} the fourth family members ($u_{4}, d_{4}, \nu_4, e_4$)  
obtain the masses at
($224$ GeV (285 GeV), $285$ GeV (224 GeV), $84$ GeV, $170$ GeV), respectively. 
For the other choice  of the next break %, more 
%trustable in our understanding, 
we could not determine the fourth family masses, but  when assuming the values for these masses we predicted mixing 
matrices in dependence on the masses. All these studies were done on the tree level. We are 
studying now symmetries of the Yukawa couplings if we go beyond the tree level. 
Let us add that the last experimental data~\cite{moscow09} from the HERA experiments require that there is  
no $d_4$ quark with the mass lower than $250$ GeV. 

Our stable  fifth family baryons, which might form the dark matter, also   
do not contradict the so far observed experimental data---as it is the measured 
(first family) baryon number and its ratio to the photon  energy density, 
as long as the fifth family quarks are 
 heavy enough ($> 1$ TeV). (This would be true for any stable heavy family.) 
Namely, all the measurements, which connect the baryon and the photon 
energy density, relate to the moment(s) in the history of 
the universe, when baryons of the first family where formed ($k_b T$ bellow the binding energy 
of the three first family quarks dressed into constituent mass of $m_{q_1}c^2 \approx 300$ MeV, that is bellow 
$10$ MeV) and the electrons and nuclei   formed  atoms ($k_b \,T 
 \approx 1$ eV). The chargeless (with respect to the colour and electromagnetic 
 charges) clusters of the fifth family were 
 formed long before (at $ k_b T\approx E_{c_5}$ (see Table~\ref{TableI.})), contributing the equal amount of the
 fifth family baryons and anti-baryons to the dark matter, provided that there is no 
 fifth family baryon---anti-baryon asymmetry   
 %%%%
 %%%%POSTAVI TO tudi V ZAKLJUCEK
 (if the asymmetry  is nonzero the colourless baryons or anti-baryons are formed also at the 
 early stage of the colour phase transition at around $1$ GeV). %(section~\ref{evolution}). 
 %%%%
 They  manifest 
 after decoupling from the plasma (with their small number density and  small cross 
 section) (almost) only their gravitational  interaction.

In this paper we estimate the properties of the fifth family members ($u_5,d_5,\nu_5,e_5$), as well 
as of the clusters of these members, in particular the fifth family neutrons, under the assumptions that:\\
%I. The fifth family neutrons and anti-neutrons are candidates to form the dark matter.\\
I. Neutron is the lightest fifth family baryon.\\ 
II. There is no fifth family baryon---anti-baryon asymmetry.\\
The assumptions are made since we are not yet able to derive the properties of the family from the starting 
Lagrange density of the approach. The results of the present paper's study are helpful 
to better understand  steps needed to come from the approach's starting Lagrange density to the 
low energy effective one.

From  the approach unifying spin and charges we  learn: \\
i. The stable fifth family members have masses higher than $\approx 1$ TeV  and smaller than  
$ \approx 10^6$ TeV. \\
ii. The stable fifth family members have the properties of the lower four families; 
that is the same family members with the same 
(electromagnetic, weak and colour) charges 
and interacting correspondingly with the same gauge fields.

We estimate the masses of the fifth family quarks by studying 
their behaviour in the evolution of the universe, their formation of %colour and electromagnetic 
chargeless 
(with respect to the electromagnetic and colour interaction) clusters and the properties of these clusters 
when scattering on the ordinary (made mostly of the first family members) matter and among themselves.  
We use a simple (the hydrogen-like) model~\cite{gnBled07} to estimate
the size and the binding energy of the fifth family baryons, assuming that the fifth family 
quarks are heavy enough to interact mostly by exchanging one gluon. 
We solve the Boltzmann equations for the fifth family quarks (and anti-quarks) 
forming the colourless clusters in the expanding universe, starting  in the energy 
region when the fifth family members are ultrarelativistic, up to $\approx 1$ GeV when the colour phase transition starts. 
In this energy interval the  one gluon exchange is the dominant interaction among quarks and the plasma. 
We conclude that the quarks and anti-quarks, which   
succeed to form neutral (colourless and electromagnetic chargeless) clusters,  have the properties 
of the dark matter constituents if their masses are within the interval of a few TeV $< m_{q_5} c^2< $ a few hundred TeV, 
while the  rest of the coloured fifth family objects annihilate within the colour phase 
transition period %starting at around $1$ GeV 
with their anti-particles for the zero fifth family baryon number asymmetry. 
%%%%
%%%%IZRACUNATI STEVILO NEVTRONOV IN ANTINEVTRONOVPRI NEKI PREDPOSTAVLJENI ASIMETRIJI, DENIMO ISTI KOT PRI PRVI DRUZINI. 
%%%%

We estimate also the behaviour of 
our fifth family clusters if hitting the DAMA/NaI---DAMA-LIBRA~\cite{rita0708} and CDMS~\cite{cdms} 
experiments presenting the limitations the DAMA/NaI experiments put on our fifth family quarks when    
%which exhibit annual modulation and 
recognizing that CDMS has not found any event (yet). 
%%%%
%%%%KOMENTIRAJ LIMITO, KI JO POSTAVI CDMS. POVEJ V POGLAVJU O DYNAMIKI, KAKO SE OBNASAJO GRUCE
%%%%TO  JE TUDI ZELJA REFEREEJA
%%%%

%%%%
%%%%POSTAVI V ZAKLJUCEK
%%%%
The fifth family baryons are not the objects (WIMPS), which would interact with only the weak interaction, 
since their decoupling from the rest of the 
plasma in the expanding universe is determined by the colour force and  
their interaction with the ordinary matter is determined with the  
fifth family "nuclear force" (this is the force among clusters of the fifth family quarks, 
manifesting much smaller cross section than does the ordinary, mostly first family, 
"nuclear force") as long as their mass is not higher than $10^{4} $ TeV, when the weak interaction starts to 
dominate as commented 
in the last paragraph of section~\ref{dynamics}.

\section{Properties of clusters of the heavy family}
\label{properties}

Let us  study the properties of the fifth family of quarks and leptons as 
predicted by the approach unifying spin and charges, with masses several orders of 
magnitude greater than those of the known three families,  decoupled in the Yukawa couplings from 
the lower mass families and with 
the charges and their couplings to the gauge fields of the known families (which all seems, 
due to our estimate predictions of the approach, reasonable assumptions).   
Families distinguish among themselves (besides in masses) 
in the family index (in the quantum number, which in the approach is determined  
by the second kind of the Clifford algebra objects' operators~\cite{pn06,n92,n93} 
$\tilde{S}^{ab}=\frac{i}{4}(\tilde{\gamma}^a \tilde{\gamma}^b - \tilde{\gamma}^b 
\tilde{\gamma}^a)$, anti-commuting with the Dirac $\gamma^a$'s), and 
(due to the Yukawa couplings)  in their masses.  

For a heavy enough family the properties of baryons (protons $p_5$ $(u_5 u_5 d_5)$, 
neutrons $n_5$ $(u_5 d_5 d_5)$, $\Delta_{5}^{-}$, $\Delta_{5}^{++}$) 
made out of  
quarks $u_5$ and $d_5$ can be estimated by using the non relativistic Bohr-like model 
with the $\frac{1}{r}$ %(radial) 
dependence of the potential  
between a pair of quarks  $V= - \frac{2}{3} \frac{\hbar c \,\alpha_c}{r}$, where $\alpha_c$ is in this case the 
colour coupling constant. 
Equivalently goes for anti-quarks. % (with the factor $\frac{4}{3}$ instead of $\frac{2}{3}$). 
This is a meaningful approximation as long as the   
one gluon exchange is the dominant contribution to the interaction among quarks,  
that is as long as excitations  of a cluster are not influenced by  the linearly rising 
part of the potential~\footnote{Let us tell that a simple bag model evaluation does not 
contradict such a simple Bohr-like model.}. The electromagnetic 
and weak interaction contributions are  of the order of $10^{-2}$ times smaller. 
%~\footnote{
%A simple bag model %~\cite{kuti}, 
%with the potential $V(r)=0 $ for $r<R$ 
%and $V(r)= \infty $ otherwise, supports our rough estimation. It, namely, 
%predicts for the lowest energy $E$ (the mass) of a cluster of three quarks:  
%$E= 3 \,m_{q_5} c^2 (1+ (x \hbar c /m_{q_5}c^2 R)^2),$ with $\tan x = x/ [1-
%(m_{q_5}c^2 R/\hbar c) - \sqrt{x^2 + (m_{q_5}c^2 R/\hbar c)^2} \,], $ where $2.04< x <\pi$ 
%for $0 < (m_{q_5}c^2 R/ \hbar c) < \infty$. For  $R$ taken from our Bohr-like model 
%$R= r_{c_5}$,  $(m_{q_5}c^2 R/ \hbar c) \approx 8$, 
%for example, is $x$ close to $3$ and rises very slowly to $\pi$. Accordingly 
%the mass of the three quark cluster is close to three masses of the quark even in the bag model, while 
% one gluon exchange suggests the binding energy for one pair of quarks of the order of 
% $\frac{1}{20} m_{q_5}c^2$}.
%
Which one of $p_5$, $n_5$, or maybe $\Delta_{5}^-$ or $\Delta_{5}^{++}$,  
is a stable fifth family baryon, depends on the ratio of the bare masses 
$m_{u_5}$ and  $m_{d_5}$, as well as on the  weak and the 
electromagnetic interactions among quarks. 
If $m_{d_5}$ is appropriately 
smaller than $m_{u_5}$ so that the  
weak and electromagnetic interactions favor the neutron $n_5$, then $n_5$ is 
a colour singlet electromagnetic chargeless stable cluster of quarks, with 
the weak charge $-1/2$. %with the lowest mass among 
%the nucleons of the fifth family. 
If $m_{d_5}$ is larger (enough, due to the stronger electromagnetic repulsion among 
the two $u_5$ than among the two $d_5$) than $m_{u_5}$, the proton $p_5$ which is 
a colour singlet stable nucleon with the weak charge $1/2$,  %and the electromagnetic charge 
needs the electron $e_5$ or $e_1$ or $\bar{p}_1$ to form  a stable  electromagnetic %(and weak) 
chargeless cluster (in the last case it could also be the weak singlet and would accordingly  manifest 
the ordinary nuclear force only).  
An atom made out of only fifth family members might be lighter or not than $n_5$, 
depending on the masses of the fifth family members. 

Neutral (with respect to the electromagnetic and colour charge) fifth family 
particles that constitute the dark matter can be $n_5,\nu_5$ 
or  charged baryons like  $p_5, \Delta^{++}_5$, $\Delta^{-}_5$, forming neutral atoms with
$e^{-}_5$ or $\bar{e}^{+}_5$, correspondingly, or (as said above) $p_{5} \bar{p}_1$ . 
We treat the case that $n_5$ as well as $\bar{n}_5$ 
form the major part of the dark matter, assuming %in this letter 
that $n_5$ (and $\bar{n}_5$) are stable baryons (anti-baryons). Taking  $m_{\nu_5}< m_{e_5}$ 
also $\nu_5$ contributes to the dark matter. We shall comment this in section~\ref{directmeasurements}.

In the Bohr-like model % or the a little more sophisticated hydrogen-like model in which 
%we let the widths of the hydrogen wave functions to be adapted as explained in Appendix~I, 
we obtain if neglecting more than one gluon exchange contribution
%(****GREGOR, %POISKATI REFERENCE IN POPRAVITI TABELO!!!!VSTAVITI v FOOTNOTE PRAVI IZRAZ, 
%Napisati Appendix****)
%
\begin{eqnarray}
\label{bohr}
E_{c_{5}}\approx -3\; \frac{1}{2}\; \left( \frac{2}{3}\, \alpha_c \right)^2\; \frac{m_{q_5}}{2} c^2,
\quad r_{c_{5}} \approx  \frac{\hbar c}{ \frac{2}{3}\;\alpha_c \frac{m_{q_5}}{2} c^2}. 
\end{eqnarray}
The mass of the cluster is approximately $m_{c_5}\, c^2 \approx  
3 m_{q_5}\, c^2(1- (\frac{1}{3}\, \alpha_c)^2)$. We use the  factor of $\frac{2}{3}$ 
for a  two quark pair potential and of $\frac{4}{3}$ for a quark and an anti-quark pair potential. %~\cite{ruhula} 
If treating correctly the three quarks' (or anti-quarks') center of mass motion in the 
hydrogen-like model, allowing  the hydrogen-like functions to adapt the width as presented in 
Appendix~I,
%\ref{betterhf},   
the factor $-3\; \frac{1}{2}\; (\frac{2}{3})^2\; \frac{1}{2}$ in Eq.~\ref{bohr} is replaced by
$0.66$, and the mass of the cluster is accordingly $3 m_{q_5} c^2(1-0.22\, \alpha_{c}^2)$, while 
the average radius takes the values as presented in  Table~\ref{TableI.}.

Assuming that the coupling constant   
of the colour charge  $\alpha_c$   runs with the kinetic energy $- E_{c_{5}}/3$ and taking into account 
the number of families which contribute to the running coupling constant in dependence on the kinetic energy 
(and correspondingly on the mass of the fifth family quarks)
%of a quark %as in %ref.~\cite{greiner}  
%
%($\,\alpha_c(E^2)=\frac{\alpha_c(M^2)}{1+\frac{\alpha_c(M^2)}{4 \pi} (11-\frac{2 N_F}{3}) 
%\textrm{ln}(\frac{E^2}{M^2}) }$, with $\alpha_{c}((91 \; \textrm{GeV})^2)=0.1176(20)$,
%the number of flavours $N_F=8$ or less, depending on the temperature (****ALI JE UPOSTEVANO?****)) 
%
we estimate  the properties of a baryon as presented on Table~\ref{TableI.} (the table 
is calculated from the hydrogen-like model presented in Appendix~I), 
\begin{table}
\begin{center}  
\begin{tabular}{||c||c|c|c|c|}
\hline
$\frac{m_{q_5} c^2}{{\rm TeV}}$ & $\alpha_c$ & $\frac{E_{c_5}}{m_{q_5} c^2}$ & 
$\frac{r_{c_5}}{10^{-6}{\rm fm}}$ & $\frac{\Delta m_{ud} c^2}{{\rm GeV}}$ 
\\
\hline
\hline
$1   $ & 0.16   & -0.016   & $3.2\, \cdot 10^3$   & 0.05		             \\
\hline
$10  $ & 0.12   & -0.009   & $4.2\, \cdot 10^2$   & 0.5           \\
\hline
$10^2$ & 0.10   & -0.006   & $52$            & 5           \\
\hline
$10^3$ & 0.08   & -0.004   & $6.0$           & 50           \\
\hline
$10^4$ & 0.07   & -0.003   & $0.7$           & $5 \cdot 10^2$           \\
\hline
$10^5$ & 0.06   & -0.003   & $0.08$          & $5 \cdot 10^3$            \\
\hline
\end{tabular}
\end{center}
\caption{\label{TableI.} 
%Na desni v vsaki deljeni celici je primer, ko je $N_F=8$, na levi pa ko je 
%$N_F=6$.
The properties of a cluster of the fifth family quarks
within the extended Bohr-like (hydrogen-like) model from Appendix~I. 
$m_{q_5}$ in TeV/c$^2$ is the assumed fifth family quark mass,
$\alpha_c$ is the coupling constant 
of the colour interaction at $E\approx (- E_{c_{5}}/3)\;$ (Eq.\ref{bohr})   
which is the kinetic energy 
of  quarks in the baryon, 
$r_{c_5}$ is the corresponding average  radius. Then  $\sigma_{c_5}=\pi r_{c_5}^2 $  
is the corresponding scattering cross section.} % for a chosen quark mass.} 
\end{table}

The binding energy is approximately  $\frac{1}{ 100}$  of the mass 
of the cluster (it is $\approx \frac{\alpha_{c}^2}{3}$).  The baryon $n_5$ ($u_{5} d_{5} d_{5}$) 
is lighter than the baryon $p_{5}$,   ($u_{q_5} d_{q_5} d_{q_5}$) 
if $\Delta m_{ud}=(m_{u_5}-m_{d_5})$ is smaller than $\approx (0.05,0.5,5, 50, 500, 5000)$ GeV  
for the six  values of the $m_{q_5} c^2$ on Table~\ref{TableI.}, respectively. %(****PREVERITI!****) 
We see  from Table~\ref{TableI.} that the ''nucleon-nucleon'' 
force among the fifth family baryons leads to many orders of 
magnitude smaller cross section than in the case 
of the first family nucleons ($\sigma_{c_5}= \pi r_{c_5}^2$ is from $10^{-5}\,{\rm fm}^2$  for 
$m_{q_5} c^2 = 1$ TeV to $10^{-14}\, {\rm fm}^2$  for $m_{q_5} c^2 = 10^5$ TeV). 
Accordingly is the scattering cross section between two  fifth family baryons   
determined by the weak interaction as soon as the mass   exceeds  several GeV.

If a cluster of the heavy (fifth family) quarks and leptons and  of the 
ordinary (the lightest) family is made, 
then, since ordinary family   dictates the radius and the excitation energies  
of a cluster, its 
properties are not far from the properties of the ordinary hadrons and atoms, except that such a  
cluster has the mass dictated by the heavy family members. 
\section{Evolution of the abundance of the fifth family members in the universe}
\label{evolution}

We assume that there is no fifth family baryon---anti-baryon asymmetry
and that the  neutron is the lightest baryon made out of the 
fifth family quarks. Under these assumptions and with the knowledge from 
our rough estimations~\cite{gmdn07} that the fifth family masses are within  the interval    
from  $1$ TeV to $10^6$ TeV  
%
% (We only have a rough estimation that this could be the case).
%%%%
%%%%
%KOMENTIRAJ NA KONCU, PRIMER, KO TA PRIVZETEK SPROSTIMO
%%%%
%%%%
we study the behaviour of our fifth family quarks and anti-quarks in the expanding  
(and accordingly cooling down~\cite{dodelson}) 
universe in the plasma of all other fields (fermionic and bosonic) from the 
period, when the fifth family members carrying all the three charges (the colour, 
weak and electromagnetic) are ultra relativistic and is their number 
(as there are the numbers of all the other fermions and bosons in the 
ultra relativistic regime) determined by the temperature. We follow the fifth family quarks and 
anti-quarks first through the 
freezing out period, when the fifth family quarks and anti-quarks 
start to have too large mass to be formed out of the plasma (due to  the plasma's too low temperature), 
then through the period when  first  the clusters of 
di-quarks and di-anti-quarks and then the colourless neutrons and anti-neutrons ($n_5$  and $\bar{n}_5$) 
are formed.  The fifth family neutrons being tightly bound into the colourless objects do not feel   
the colour phase transition  when it  
starts bellow $ k_b  T\approx 1$ GeV ($k_b$ is the Boltzmann constant) and %$n_5$  and $\bar{n}_5$
decouple accordingly from the rest of quarks and anti-quarks and gluons and manifest 
today as the dark matter constituents. 
We take the quark mass as a free parameter in the interval 
from $1$ TeV to $10^6$ TeV and determine the mass from the observed dark matter density.  
%assuming that there are $n_5$ and 
%$\bar{n}_5$ which form the today's dark matter.  

At the colour phase transition, however, the coloured fifth family quarks and anti-quarks 
annihilate  to the today's unmeasurable density: Heaving much larger mass (of the order of $10^{5} $ times larger),  
%as we shall evaluate from the dark matter density for our fifth family quarks)  and accordingly 
and correspondingly much larger momentum (of the order of $10^3$ times larger) as well as much 
larger binding energy (of the order 
of $10^5$ times larger)
than the first family quarks when they are  "dressed" into constituent mass,   
the coloured fifth family quarks succeed in the colour phase transition region  to annihilate with the 
corresponding anti-quarks to the non measurable extend, 
if it is no fifth family baryon asymmetry. 
%%%%
%%%%
%NAPRAVI RACUN
%KOMENTIRAJ, CE ASIMETRIJA JE, TEDAJ PAC PRISPEVAJO K TEMNI MASI. TUDI TEDAJ JE NJIHOVA MASA 
% DOLOCENA S TEM, DA MORAJO BITI DOVOLJ TEZKI, KER SICER IYJAVA NE VELJA.
%%%%
%%%%

In the freezing out period almost up to the colour phase transition
the kinetic energy of quarks is high enough so that  
the one gluon exchange 
dominates in the colour interaction of quarks with the plasma, while the (hundred times) weaker weak 
%interaction and even much weaker 
and electromagnetic 
interaction can be neglected. 

The quarks and anti-quarks start to freeze out when the temperature of the plasma falls close to  
$m_{q_5}\,c^2/k_b $. They are forming clusters (bound states) 
when the temperature falls close to  the binding energy (which is due to Table~\ref{TableI.} 
$\approx \frac{1}{100} m_{q_5} c^2$). 
When the three quarks (or three anti-quarks) of the 
fifth family form a colourless baryon (or anti-baryon), they decouple from the rest of plasma due to small 
scattering cross section manifested by the average radius presented in Table~\ref{TableI.}.

%To estimate the behaviour of our stable heavy family quarks and anti-quarks in the expanding 
%universe we would need to know the mass of the fifth family quarks and the fifth family baryon asymmetry. 
%%We take the quark mass as a free parameter and determine it from the observed dark matter density, 
%assuming that there are $n_5$ and 
%$\bar{n}_5$ which form the today's dark matter.  

Recognizing that at the temperatures ($ 10^6$ TeV $> k_b T >1 $ GeV) the one gluon exchange gives the 
dominant contribution to the interaction among quarks of any family, it is not difficult to estimate 
the thermally averaged scattering cross sections (as the function of the temperature) for 
the fifth family quarks and   anti-quarks to 
scatter:\\    
$\;\;\;\;$ i. into all the 
relativistic quarks and anti-quarks of lower mass families ($<\sigma v>_{q\bar{q}}$),\\  
$\;\;\;$ ii.  into gluons ($<\sigma v>_{gg}$), \\
$\;\;$ iii.  into (annihilating) bound states of a fifth family quark and an anti-quark mesons $\;\;\;$
($<\sigma v>_{(q\bar{q})_b}$), \\ 
%with the summation over $2 \times 8$ gluons) 
$\;\;$ iv. into bound states of two fifth family quarks and into the fifth family baryons 
($<\sigma v>_{c_5}$)  %which here 
%we evaluate only with the assumptions that these two cross section can be treated together  
(and equivalently into two anti-quarks and 
into anti-baryons).

%%%%
%%%%
%KAR SLEDI JE ZE BILO POVEDANO IN NE PONAVLJAJ
%%%%
%%%%
%A rough estimation of the probability for the first family quarks and anti-quarks   to annihilate  
%at the colour phase transition ($  k_b T\approx 1$ GeV) shows that the fifth family quarks do 
%annihilate with the fifth family anti-quarks, leaving (almost, negligible small) number of stable 
%mesons of the first and the fifth family members. 
%%iv.) the ''Thompson'' scattering cross section for gluons on quarks or anti-quarks ($\sigma_{T}$) 

The one gluon exchange scattering cross sections are namely (up to the strength of the  coupling constants 
and up to the numbers of the order one 
determined by the corresponding groups) 
equivalent to the corresponding cross sections for the one photon exchange scattering cross sections, and  
we use correspondingly also the  expression  for scattering of an electron and a proton into 
the bound  state of a hydrogen when treating the scattering of two quarks into the bound states.
%So we use the 
%adapted cross sections from the electromagnetic interactions%when there are no  expressions available 
%in the literature for the one gluon exchange
We take the roughness  
of such estimations into account by two parameters: 
The parameter $\eta_{c_5}$ takes care of scattering 
of two quarks (anti-quarks) into three 
colourless quarks (or anti-quarks), which are the fifth family baryons (anti-baryons) 
and about the uncertainty with which this 
cross section is estimated. $\eta_{(q \bar{q})_b}$ takes care of the roughness of the used formula 
for $<\sigma v>_{(q\bar{q})_b}$.

%$\eta_{c_5}$ and $\eta_{(q \bar{q})_b}$ (defined bellow), which 
%define accordingly the accuracy with which the fifth family mass is estimated and correspondingly 
%the acceptable fifth family mass interval. 

The following expressions for the thermally averaged cross sections are used%in the Boltzmann equations 
\begin{eqnarray}
\label{sigmasq}
< \sigma v>_{q\bar{q}} &=&  \frac{16 \,\pi}{9} 
\;\left( \frac{\alpha_{c} \hbar c}{m_{q_{5}}\,c^2}\right)^2 \, c ,\nonumber\\
< \sigma v>_{gg} &=&  \frac{37 \,\pi}{108}\;\left( \frac{\alpha_{c} 
\hbar c}{m_{q_{5}}\,c^2}\right)^2\, c, \nonumber\\
< \sigma v>_{c_5} &=& \eta_{c_5}\; 10 \;\left( \frac{\alpha_{c} \hbar c}{m_{q_5}\,c^2} \right)^2\, c\; 
\sqrt{\frac{ E_{c_5}}{ k_b T}} \ln{\frac{E_{c_5}}{ k_b T}}, \nonumber\\
<\sigma  v>_{(q \bar{q})_b}&=& \eta_{(q \bar{q})_b} \;10 \;\left(
\frac{\alpha_{c} \hbar c}{m_{q_5}\,c^2}\right)^2\, c\; 
\sqrt{\frac{ E_{c_5}}{ k_b T}} \ln{\frac{E_{c_5}}{ k_b T}}, \nonumber\\
\sigma_{T } &=&  \frac{8 \pi}{3} \left(\frac{\alpha_{c} \hbar c }{m_{q_5} \, c^2}\right)^2,
\end{eqnarray}
where $v$ is the relative velocity between the fifth family  quark and its anti-quark, or between two quarks and 
$E_{c_5}$ is the binding energy for a cluster (Eq.~\ref{bohr}). 
%$< \sigma v> \;$ is the thermally averaged scattering cross section 
%times the  relative velocity: 
%i. $< \sigma v>_{q\bar{q}} $ for all the pairs  
%of the fifth family  quarks and anti-quarks into all the lower mass (of the four families') 
%quarks and anti-quarks, which are, while scattering takes place, ultra relativistic. 
%ii. $< \sigma v>_{gg}$  for scattering into gluons.  
%iii. $ <\sigma v>_{c_5}$ for two quarks (or two anti-quarks) to scatter into a bound state of 
%two quarks (anti-quarks) and from two to three quarks (anti-quarks) colourless clusters. 
%We use the equivalent expression as for scattering of an electron and a proton into 
%the bound  state of a hydrogen. 
%The parameter $\eta_{c_5}$ takes care of scattering 
%of two quarks (anti-quarks) into three 
%colourless quarks (or anti-quarks), which are the fifth family baryons (anti-baryons) and about the uncertainty with which this 
%cross section is estimated. 
%
%
%iv. $<\sigma v>_{(q\bar{q})_b}$ for  scattering into a bound state of the fifth family quark and anti-quark,  
%which annihilate in the time $\tau_{(q\bar{q})_b} < 10^{-28}$ s. 
%$\eta_{(q \bar{q})_b}$ takes care of the roughness of the used formula. 
%
$\sigma_{T }$ is the Thompson-like scattering cross section of gluons on quarks (or anti-quarks).  
%responsible for destroying the bound states of baryons. %  an acceptable approximation, when 
%the $\hbar \omega$ of a gluon is much smaller than the $m_{q_5} c^2$ and high enough that it scatters 
%elastically. 
%%%%%

To see how many fifth family quarks and anti-quarks of a chosen mass  form the fifth family 
baryons and anti-baryons today we solve the coupled systems of 
Boltzmann equations presented bellow as  a function of time (or temperature). 
The value of the fifth family quark mass which predicts the today observed dark matter is 
 the mass we are looking for. Due to the inaccuracy of the estimated 
scattering cross sections 
entering into the Boltzmann equations we %only can 
tell the interval within which the mass lies. 
We follow in our derivation of the Boltzmann equations (as much as possible) 
%, when estimating the number density of the fifth family quarks $n_{q_5}$ and 
%anti-quarks $n_{\bar{q}_5}$ clustered into baryons (with the number density $n_{c_5}$) and anti-baryons 
%($n_{\bar{c}_5}$), which to our prediction form  the dark matter today, 
the ref.~\cite{dodelson}, chapter 3. 
%$n_{q_{5}}$ is the number density of all the  fifth family quarks of any colour and spin and 
% correspondingly is assumed for the other number densities. 

Let $T_0$ be the today's black body radiation temperature, $T(t)$ the actual (studied) temperature, 
$a^2(T^0) =1$ and $a^2(T)= a^2(T(t))$ is the metric tensor component in 
the expanding flat universe---the Friedmann-Robertson-Walker metric:  
${\rm diag}\,  g_{\mu \nu} = 
(1, - a(t)^2, - a(t)^2, - a(t)^2),\;$  $(\frac{\dot{a}}{a})^2= \frac{8 \pi G}{3} \rho$, 
with $\rho= \frac{\pi^2}{15} \, g^*\, T^4$, $ \, T=T(t)$,  
$g^*$  measures the number of degrees of freedom of those of the  
four family members (f) and  gauge bosons (b), which are at the treated temperature $T$ 
ultra-relativistic ($g^*= \sum_{i\in {\rm b}} \,g_i + \frac{7}{8} \sum_{i\in {\rm f}} \,g_i$). 
$H_0 \, \approx 1.5\,\cdot 10^{-42} \,\frac{{\rm GeV} c}{\hbar c} $ is the present Hubble constant 
and $G = \frac{\hbar c }{ (m_{pl}^2)}$, $m_{pl} c^2 =  
1.2 \cdot 10^{19}$ GeV. 

Let us write down the Boltzmann equation, which treats in the expanding universe 
the number density of all the fifth 
family quarks as a function of time $t$. The fifth family quarks scatter with  anti-quarks into 
all the other relativistic quarks (with the number density $n_{q}$) and anti-quarks ($n_{\bar{q}}$ 
($< \sigma v>_{q\bar{q}}$) and into gluons 
($< \sigma v>_{gg}$). At the beginning, when the quarks are becoming non-relativistic and   
start to freeze out, the  formation of bound states is negligible. One finds~\cite{dodelson} 
the Boltzmann equation for the fifth family quarks $n_{q_5}$ (and  equivalently  for 
anti-quarks $n_{\bar{q}_5}$)
\begin{eqnarray}
\label{boltzq1}
a^{-3}\frac{d( a^3 n_{q_5})}{dt} &=& < \sigma v>_{q\bar{q}}\; n^{(0)}_{q_5} n^{(0)}_{\bar{q}_5}\,
\left( - \frac{n_{q_5} n_{\bar{q}_5}}{n^{(0)}_{q_5} n^{(0)}_{\bar{q}_5}} + 
 \frac{n_{q} n_{\bar{q}}}{n^{(0)}_{q} n^{(0)}_{\bar{q}}} \right) + \nonumber\\ 
&&< \sigma v>_{gg} \; 
n^{(0)}_{q_5} n^{(0)}_{\bar{q}_5}\,
\left( - \frac{n_{q_5} n_{\bar{q}_5}}{n^{(0)}_{q_5} n^{(0)}_{\bar{q}_5}} +  
\frac{n_{g} n_{g}}{n^{(0)}_{g} n^{(0)}_{g}} \right). 
\end{eqnarray}
Let us tell that $n^{(0)}_{i} = g_i\, (\frac{m_i c^2  k_b T}{(\hbar c)^2})^{\frac{3}{2}} 
e^{-\frac{m_i c^2}{ k_b T}}$ for 
$m_i c^2 >>  k_b T$  %(which is our case 
and   $\frac{g_i}{\pi^2}\, (\frac{ k_b T}{\hbar c})^3$ 
for $m_i c^2 << k_b T$.
Since the ultra-relativistic quarks and anti-quarks of the lower families are in the  
thermal equilibrium with the plasma and so 
are gluons, it follows $\frac{n_{q} n_{\bar{q}}}{n^{(0)}_{q} n^{(0)}_{\bar{q}}}=1= 
\frac{n_{g} n_{g}}{n^{(0)}_{g} n^{(0)}_{g}}$. Taking into account that $(a\, T)^3 \, g^*(T)$ is a constant  
it is appropriate~\cite{dodelson} to introduce a new parameter $x=\frac{m_{q_5}c^2}{k_b T}$  and 
the quantity $Y_{q_5}= 
n_{q_5}\, (\frac{\hbar c}{k_b T})^3$, $Y^{(0)}_{q_5}= 
n^{(0)}_{q_5}\, (\frac{\hbar c}{k_b T})^3$.  When taking into account that  the number of 
quarks is the same as the number of anti-quarks, and that 
$\frac{dx}{dt} = \frac{h_m \,m_{q_5}c^2}{x} $, with $h_m = \sqrt{\frac{4 \pi^3 g^*}{45}}\, 
\frac{c}{\hbar c \, m_{pl} c^2}$, Eq.~\ref{boltzq1} transforms into $\frac{dY_{q_5}}{dx} = 
\frac{\lambda_{q_5}}{x^2}\, (Y^{(0)2}_{q_5} - Y^{2}_{q_5}), $ with $\lambda_{q_5} = \frac{(<\sigma v>_{q\bar{q}} + 
<\sigma v>_{gg}) \, m_{q_5} c^2}{h_{m}\, (\hbar c)^3}$. It is this equation which  we are solving 
(up to the region of $x$ when the clusters of quarks 
and anti-quarks start to be formed) to see 
the behaviour of the fifth family quarks as a function of the temperature.  

When the temperature of the expanding universe falls close enough to the binding energy of the cluster 
of the fifth family quarks (and anti-quarks), the bound states of quarks (and anti-quarks) and the 
clusters of fifth family baryons (in our case neutrons $n_{5}$) (and anti-baryons $\bar{n}_{5}$---anti-neutrons) 
start to form. 
%%%%%
%%%%%
To a fifth family di-quark  ($q_5 + q_5 \rightarrow $ di-quark + gluon) 
a third quark clusters ( di-quark $+ q_5 \rightarrow c_5 +$ gluon) to form the colourless fifth 
family neutron (anti-neutron), in an excited state (contributing gluons back into the plasma in the thermal bath 
when going into the ground state), all in thermal equilibrium. Similarly goes with the anti-quarks clusters.
We take into account both processes approximately within the same 
equation of motion by correcting the averaged amplitude $< \sigma v>_{c_5} $ for quarks to scatter into 
a bound state of di-quarks with the parameter $\eta_{c_5}$, 
as explained above.  
%%%%%
%%%%%
The corresponding Boltzmann equation for the number of baryons $n_{c_5}$ then reads
\begin{eqnarray}
\label{boltzc}
a^{-3}\frac{d( a^3 n_{c_5})}{dt} &=& < \sigma v>_{c_5}\; n^{(0)^2}_{q_5}\,
\left( \left( \frac{n_{q_5}}{n^{(0)}_{q_5}} \right)^2 -  
\frac{n_{c_5}}{ n^{(0)}_{c_5}} \right).
\end{eqnarray}
Introducing  again $Y_{c_5}= n_{c_5}\, (\frac{\hbar c}{k_b T})^3$, 
$Y^{(0)}_{c_5}= n^{(0)}_{c_5}\, (\frac{\hbar c}{k_b T})^3$ and 
$\lambda_{c_5} = \frac{<\sigma v>_{c_5}  \, m_{q_5} c^2}{h_m\, (\hbar c)^3}$, with the 
same $x$ and $h_m$ as above, we obtain the equation 
 $\frac{dY_{c_5}}{dx} = 
\frac{\lambda_{c_5}}{x^2}\, (Y^{2}_{q_5} - Y_{c_5} \,Y^{(0)}_{q_5}\, 
\frac{Y^{(0)}_{q_5}}{Y^{(0)}_{c_5}} )$. 

The number density of the fifth family quarks $n_{q_5}$  (%and anti-quarks %
and correspondingly  $Y_{q_5}$), which has above the 
temperature of the binding energy of the clusters of the fifth family quarks (almost) 
reached the decoupled value, starts to decrease again due  to the formation of the clusters of the 
fifth family quarks (and anti-quarks) as well as due to forming the bound state of 
the fifth family quark with an anti-quark, which annihilates into gluons.  
It follows  
\begin{eqnarray}
\label{boltzq2}
a^{-3}\frac{d( a^3 n_{q_5})}{dt} &=& 
< \sigma v>_{c_5}\; n^{(0)}_{q_5}\, n^{(0)}_{q_5} 
\left[ -\left( \frac{n_{q_5}}{n^{(0)}_{q_5}} \right)^2 + \frac{n_{c_5}}{ n^{(0)}_{c_5}} -  
\frac{\eta_{(q\bar{q})_b}}{\eta_{c_5}} \;\left( \frac{n_{q_5}}{n^{(0)}_{q_5}} \right)^2 \right] + \nonumber\\ 
&& < \sigma v>_{q\bar{q}}\; n^{(0)}_{q_5} n^{(0)}_{\bar{q}_5}\,
\left(- \frac{n_{q_5} n_{\bar{q}_5}}{n^{(0)}_{q_5} n^{(0)}_{\bar{q}_5}} + 
 \frac{n_{q} n_{\bar{q}}}{n^{(0)}_{q} n^{(0)}_{\bar{q}}} \right) + \nonumber\\ 
&&< \sigma v>_{gg} \; 
n^{(0)}_{q_5} n^{(0)}_{\bar{q}_5}\,
\left(- \frac{n_{q_5} n_{\bar{q}_5}}{n^{(0)}_{q_5} n^{(0)}_{\bar{q}_5}} +  
\frac{n_{g} n_{g}}{n^{(0)}_{g} n^{(0)}_{g}} \right), 
\end{eqnarray}
with $\eta_{(q\bar{q})_b}$ and $\eta_{c_5}$ defined in Eq.~\ref{sigmasq}.
Introducing the above defined $Y_{q_5}$  and $Y_{c_5}$ the Eq.~\ref{boltzq2} transforms into 
$\frac{dY_{q_5}}{dx} = 
\frac{\lambda_{c_5}}{x^2}\, (- Y^{2}_{q_5} + Y_{c_5} \,Y^{(0)}_{q_5}\, 
\frac{Y^{(0)}_{q_5}}{Y^{(0)}_{c_5}} ) + \frac{\lambda_{(q\bar{q})_b}}{x^2}\, (-  Y^{2}_{q_5})
+ \frac{\lambda_{q_5}}{x^2}\, (Y^{(0)2}_{q_5} - Y^{2}_{q_5})$, 
with $\lambda_{(q\bar{q})_b} = \frac{<\sigma v>_{(q\bar{q})_b}  \, m_{q_5} c^2}{h_m\, (\hbar c)^3}$ 
(and with the same $x$ and $h_m$ as well as $\lambda_{c_5}$ and $\lambda_{q_5}$
as defined above). We solve this equation together with the above equation 
for $Y_{c_5} $. 

%Let us look also at the Thompson scattering of gluons on the bound states, 
%destroying clusters, which 
%starts to be negligible when the rate for gluons to scatter off the quarks ($n_{q_5} \,\sigma_T \,c$) 
%starts to be smaller than the expansion rate ($H= \sqrt{\frac{8\pi^3 \,g^*}{45}} 
%\; \frac{( k_b T)^2 \,c}{\hbar c\, m_{pl} c^2}$), with $g^*$ defined above. 
%Recognizing that the binding energy of Table~\ref{TableI.} is approximately  $\frac{1}{100}$ the mass 
%of the fifth family quarks we get the requirement that the bound states can be formed when 
%$n_{q_5} << 3. 10^{-25} (\frac{m_{q_5} \, c^2}{{\rm GeV}})^4$ 
%fm$^{-3}$, which for $m_{q_5}\, c^2 = 1$ TeV gives $n_{q_5} << 3. 10^{-13}$ fm$^{-3}$ and  for 
%$m_{q_5}\, c^2 = 10$ TeV gives $n_{q_5} << 3. 10^{-9}$ fm$^{-3}$. One can easily check from the 
%solutions of the Boltzmann equations that this requirements are fulfilled. %(****GREGOR, PREVERITE*****)

\begin{figure}[h]
\begin{center}
\includegraphics[width=15cm,angle=0]{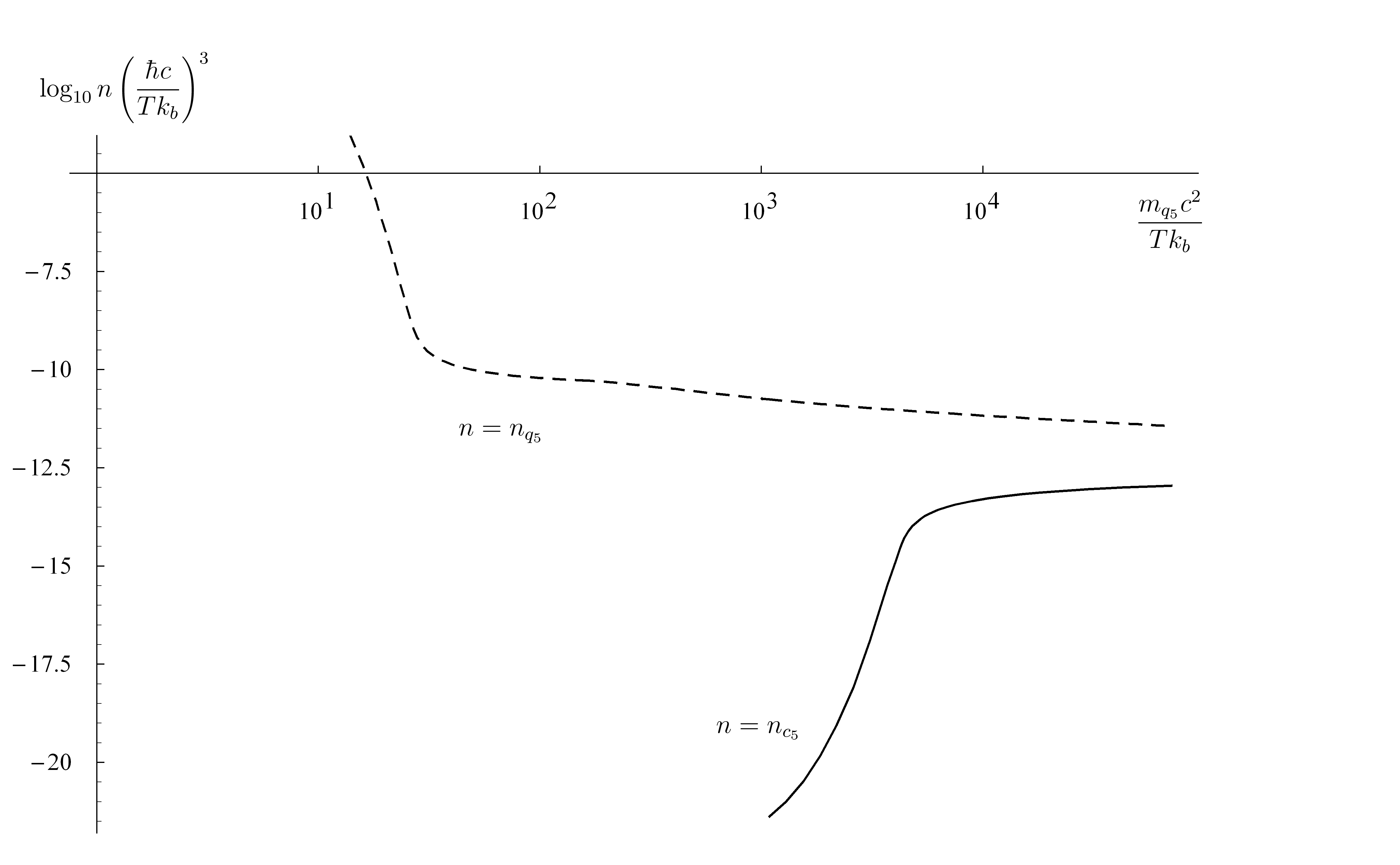}
\caption{The dependence of the two number densities $n_{q_5}$ (of the fifth family quarks) and $n_{c_5}$ (of 
the fifth family clusters) as function of $\frac{m_{q_5} \, c^2}{ k_b \, T}$ is presented 
for the special values $m_{q_5} c^2= 71 \,{\rm TeV}$, $\eta_{c_5} = \frac{1}{50}$ and $\eta_{(q\bar{q})_b}=1$.
We take $g^*=91.5$. In the treated energy (temperature $ k_b T$) interval the one gluon exchange gives the  main 
contribution to the scattering cross sections of Eq.(\ref{sigmasq}) entering into the Boltzmann equations for 
$n_{q_5}$ and $n_{c_5}$. In the figure we make a choice of the  parameters within the estimated  intervals.}
\end{center}
\label{DiagramI.}
\end{figure}

Solving the  Boltzmann  equations (Eqs.~\ref{boltzq1},~\ref{boltzc},~\ref{boltzq2}) we obtain 
 the number density of the fifth family quarks  $n_{q_5}$ (and 
anti-quarks) and the number density of the fifth family baryons $n_{c_5}$ (and anti-baryons)
 as a function of the parameter $x=\frac{m_{q_5} c^2}{k_b T}$   and the two parameters $\eta_{c_5}$ 
 and $\eta_{(q\bar{q})_b}$. The evaluations are made, as we explained above, 
with the approximate expressions for the thermally averaged cross sections  from Eq.(~\ref{sigmasq}), 
corrected  by the parameters  
$\eta_{c_5}$  and  $\eta_{(q \bar{q})_b}$ (Eq.~\ref{sigmasq}). 
%%%%%%%%%%%%%%%%%%
%%%%%%%%%%%%%%%%%
We made a rough estimation of the two intervals, within which the parameters 
$\eta_{c_5}$  and  $\eta_{(q \bar{q})_b}$ (Eq.~\ref{sigmasq}) seem to be acceptable. 
More accurate evaluations 
of the cross sections are under consideration.
%%%%%%%%%%%%%%%
%%%%%%%%%%%%%%%
In fig.~\ref{DiagramI.} both number densities (multiplied by $(\frac{\hbar \, c}{ k_b T})^3$, which is 
$Y_{q_5}$ and $Y_{c_5}$, respectively for the quarks and the clusters of quarks) as a function 
of  $ \frac{m_{q_5} \, c^2}{ k_b T}$ for $\eta_{(q\bar{q})_3}=1$ and $\eta_{c_5}=\frac{1}{50}$ are presented. 
%%%%%%%%%%%%%%
%%%%%%%%%%%%%%%%
The particular choice of the parameters $\eta_{(q\bar{q})_3}$ and $\eta_{c_5}$ in fig.~\ref{DiagramI.}
is made as a typical example.
%%%%%%%%%%%%%%%%%%%
%%%%%%%%%%%%%
The calculation is performed up to $ k_b T=1$ GeV (when the colour phase transition starts and the one 
gluon exchange stops to be the acceptable approximation).

Let us repeat how the $n_5$ and $\bar{n}_5
$ evolve in the evolution of our universe. 
 The quarks and anti-quarks are at high temperature 
($\frac{m_{q_5} c^2}{k_b T}<< 1$) in thermal equilibrium with the plasma (as are also all the other 
families and bosons of  lower masses). 
As the temperature of the plasma (due to the expansion of the universe) drops close to the 
mass of the fifth family 
quarks, quarks and anti-quarks scatter into all the other (ultra) relativistic fermions and bosons, but 
can not be created any longer from the plasma (in the average).  
At the temperature close to  the binding energy of the quarks in a cluster, the clusters of  the fifth family 
($n_{c_5}, n_{\bar{c}_5}$)
baryons start to be formed. We evaluated the number density  $n_{q_5} (T) \,
(\frac{\hbar c}{ k_b T})^3 = Y_{q_5} $  
of the fifth family quarks (and anti-quarks) and the number density of the fifth family baryons 
$n_{c_5} (T) \,(\frac{\hbar c}{ k_b T})^3 = Y_{c_5} $ for several choices  of  
$m_{q_5}, \eta_{c_5}$ and $\eta_{(q \bar{q})_b}$ up to $ k_b T_{lim}= 1$ GeV $=\frac{m_{q_5} c^2}{x_{lim}} $.  
%The fifth family quark mass then follows from Eq.~\ref{dm}. 

From the calculated decoupled number density of baryons and anti-baryons of the fifth family quarks  
(and anti-quarks) $n_{c_5}(T_1)$ at temperature $ k_b T_1=1$ GeV, where we stopped our 
calculations  as a function of the quark mass and of the 
two parameters $\eta_{c_5}$ and $\eta_{(q\bar{q})_b}$, the today's mass density of the dark matter 
follows (after taking into account that when once the $n_{5}$ and $\bar{n}_{5}$ decouple, their number 
stays unchanged but due to the expansion of the universe their density decreases according to 
$a^{3}_1 n_{c_5}(T_1)= 
a^{3}_2 n_{c_5}(T_2)$, with the today's $a_0=1$ and the temperature $T_0=2.725^0 $ K) leading to~\cite{dodelson}
\begin{eqnarray}
\label{dm}
\rho_{dm} &=& \Omega_{dm} \rho_{cr}= 2 \, m_{c_5}\, n_{c_5}(T_1) \, 
\left(\frac{T_0}{T_1} \right)^3 \frac{g^*(T_1)}{g^*(T_0)},
\end{eqnarray}
where we take into account that $g^*(T_1) (a_1 T_1)^3= g^*(T_0)(a_0 T_0)^3$, 
with $T_0 = 2.5 \,\cdot 10^{-4}\,\frac{{\rm eV}}{k_b}$, $g^*(T_0)= 2 + \frac{7}{8}\,\cdot3 \,\cdot \,
(\frac{4}{11})^{4/3}$, $g^*(T_1)= 2 + 2\,\cdot 8 + \frac{7}{8}\, (5\cdot3\cdot2\cdot2 + 6\cdot 
2\cdot2)$ and  
 $\rho_{cr} \,c^2 \,\approx \frac{3\, H^{2}_{0}\, c^2}{8 \pi G} %\approx 7\cdot 10^{-45} {\rm GeV} (fm)^{-3}
 \approx 5.7\, \cdot 10^3 \frac{{\rm eV}}{{\rm cm}^3}$, factor $2$ counts 
 baryons and anti-baryons. % (since the spin  of baryons is taken into account in $n_{c_5}$).  

The  intervals for the acceptable  parameters $\eta_{c_5}$ and $\eta_{(q \bar{q})_b}$  (determining  
the inaccuracy, with which the scattering cross sections were evaluated) influence 
the value of $n_{c_5}$ and  
determine the interval, within which one expects the fifth family mass.
\begin{table}
\begin{center} \begin{tabular}{|c||c|c|c|c|c|}
\hline
$\frac{m_{q_5} c^2}{{\rm TeV}}$&$\eta_{(q\bar{q})_b}=\frac{1}{10}$&$\eta_{(q\bar{q})_b}=\frac{1}{3}$&$\eta_{(q\bar{q})_b}=1$& 
$\eta_{(q\bar{q})_b}=3$&$\eta_{(q\bar{q})_b}=10$\\
\hline\hline
$\eta_{c_5}=\frac{1}{50}$ & 21 & 36 & 71& 159&417\\
\hline
$\eta_{c_5}=\frac{1}{10} $ & 12 & 20 & 39&  84&215 \\
\hline
$\eta_{c_5}=\frac{1}{3} $ & 9  & 14 & 25&  54&134 \\
\hline
$\eta_{c_5}=1 $           & 8  & 11 & 19&  37 &88 \\
\hline
$\eta_{c_5}= 3$ &           7  & 10 & 15&  27 &60 \\
\hline
$\eta_{c_5}=10$ &           7* &  8*& 13&  22 &43 \\
\hline
\end{tabular}
\end{center}
\caption{\label{TableII.} The fifth family quark mass is presented (Eq.(\ref{dm})), calculated for different 
choices of $\eta_{c_5}$ (which takes care of the inaccuracy with which  a colourless cluster of three 
quarks (anti-quarks) cross section was estimated  and of $\eta_{(q\bar{q})_b}$ (which takes care of the inaccuracy
with which the cross section for the 
annihilation of a bound state of quark---anti-quark was taken into account) from Eqs.~(\ref{dm}, \ref{boltzc}, \ref{boltzq1}).
* denotes non stable calculations.}
\end{table}
We read from Table~\ref{TableII.} the mass interval for the fifth family quarks' mass, 
which fits Eqs.~(\ref{dm}, \ref{boltzc}, \ref{boltzq1}):
\begin{eqnarray}
\label{massinterval}
10 \;\; {\rm TeV} < m_{q_5}\, c^2 < {\rm a\, few} \cdot 10^2 {\rm TeV}.
\end{eqnarray}
From this mass interval we estimate from Table~\ref{TableI.} the cross section for the 
fifth family neutrons $\pi (r_{c_5})^2$:
\begin{eqnarray}
\label{sigma}
10^{-8} {\rm fm}^2 \, < \sigma_{c_5} < \, 10^{-6} {\rm fm}^2.
\end{eqnarray}
(It is  at least $10^{-6} $ smaller than the cross section for the first family neutrons.)

Let us comment on the fifth family quark---anti-quark annihilation at the colour phase transition, 
which starts at 
approximately $1$ GeV. When the colour phase transition starts, the quarks start to "dress" into 
constituent mass, which brings to them 
 $\approx 300$ MeV/$c^2$, since to the force many gluon exchanges start to contribute. 
 The scattering cross sections, which were up to 
 the phase transition dominated by one gluon exchange, rise now to the value of a few ${\rm fm}^2$ and more,
 say $(50 {\rm fm})^2$. 
 Although the colour phase transition is not yet 
 well understood even for the first family quarks, the evaluation of what happens to the 
 fifth family quarks and anti-quarks and coloured clusters of the fifth family 
 quarks or anti-quarks can still be done as follows. 
 At the interval, when the temperature $ k_b T$ is 
considerably above the binding energy of the "dressed" first family quarks and anti-quarks  
into mesons or of the binding energy of the 
three first family quarks 
or anti-quarks into the first family baryons or anti-baryons, which is $\approx $ a few MeV 
(one must be more careful with the mesons), 
the first family quarks  and anti-quarks 
move in the plasma like being free. (Let us remind the reader 
that the nuclear interaction can be derived as the interaction among the  clusters of quarks~\cite{bmmn}.) 
25 years ago there were several proposals to treat nuclei as  clusters of dressed 
quarks instead of as clusters of baryons. Although this idea was not very fruitful (since even models 
with nuclei as  bound states of $\alpha$ 
particles work many a time reasonably) it also was not far from the reality. 
Accordingly it is meaningful to accept the description 
of plasma at temperatures above a few $ 10$ MeV/$k_b$ as the plasma of less or more "dressed" quarks 
%(where "dressing" matters only the 
%first family quarks)   
with the very large scattering amplitude (of   $\approx (50{\rm fm})^2$). 
The fifth family quarks and anti-quarks, heaving much higher mass (several ten  thousands 
GeV/$c^2$ to be compared with  $\approx 300$ MeV/$c^2$) than the first family quarks and accordingly 
much higher momentum,  "see" the first family quarks as  a "medium"  in which they (the fifth family
quarks) scatter among themselves. The fifth family quarks and anti-quarks, 
having much higher binding energy when forming a meson among themselves  than when forming  
mesons with the first family 
quarks and anti-quarks (few thousand GeV  to be compared with few MeV or few $10$ MeV)  
and correspondingly very high annihilation probability and 
also pretty low velocities ($\approx 10^{-3} c$),  have during the scattering 
enough time to annihilate with their anti-particles.
The ratio of the scattering time between two coloured quarks (of any kind) and the Hubble time 
is of the order of $\approx 10^{-18}$ and therefore although the number of the fifth family 
quarks and anti-quarks is  of the order of $10^{-13}$ smaller than the number of the quarks and anti-quarks 
of the first family (as show the solutions of the Boltzmann equations 
presented in fig.~\ref{DiagramI.}),  the fifth family quarks and anti-quarks have in the first period 
of the colour phase transition 
(from $\approx$ GeV to $\approx 10$ MeV) enough opportunity to scatter often enough among themselves to deplete (their  
annihilation time is for several orders of magnitude smaller than
  the time needed to pass by). 
 More detailed calculations, which are certainly needed, are under considerations. Let us still do 
 rough estimation about the number of the coloured fifth family quarks (and anti-quarks).
 Using the expression for the thermally averaged cross section for scattering of a  quark and an 
 anti-quark and annihilating  ($<\sigma  v>_{(q \bar{q})_b} $ from Eq.(\ref{sigmasq})) 
 and correcting the  part which determines 
 the scattering cross section by replacing it with $ \eta\, ( 50 {\rm fm})^2 c\; $ (which takes into account the scattering in the 
 plasma during the colour phase transition in the expanding universe) we obtain the  expression 
 $ <\sigma  v>_{(q \bar{q})_b}= \eta_{(q \bar{q})_b} \, \eta\, ( 50 {\rm fm})^2 c\; 
\sqrt{\frac{ E_{c_5}}{ k_b T}} \ln{\frac{E_{c_5}}{ k_b T}}$, which is almost independent of the velocity of 
the fifth family 
quarks (which slow down when the temperature lowers). We shall assume that the temperature is 
lowering as it would be no phase transition and correct this fact with the parameter $\eta$, which could  
for a few orders of magnitude (say $10^2$) enlarge the depleting probability.  
Using this expression for $<\sigma  v>_{(q \bar{q})_b}$ in 
the expression for $\lambda= \frac{<\sigma  v>_{(q \bar{q})_b} \; m_{q_5} c^2}{h_m (\hbar c)^3}$, we 
obtain for a factor up to $10^{19}$ larger $\lambda$ than it was the one dictating the freeze out 
procedure of $q_5$ and $\bar{q}_{5}$ before the phase transition. 
Using then the equation $\frac{dY_{q_5}}{dx} = 
\frac{\lambda_{c_5}}{x^2}\, (- Y^{2}_{q_5})$ and integrating it from $Y_1$ which is the value 
from the fig.~\ref{DiagramI.}
at $1$ GeV up to the value when $ k_b T\approx 20 $ MeV, when the first family quarks start to bindd into baryons, 
we obtain in the approximation that $\lambda $ is independent of $x$ (which is not really the case) that 
$\frac{1}{Y(20 {\rm MeV})}= 10^{32} \frac{1}{2\cdot 10^5} $ or $Y(20 {\rm MeV}) = 10^{-27}$ and correspondingly
$n_{q_5}(T_0)= \eta^{-1} 10^{-24} cm^{-3}$. Some of these fifth family quarks can form the mesons or baryons and anti-baryons 
with the first family quarks $q_1$ when they start to form baryons and mesons. They would form the 
anomalous hydrogen in the ratio: $\frac{n_{ah}}{n_{h}} \approx  \eta^{-1} \cdot 10^{-12} $, where 
$n_{ah}$ determines the number of the anomalous (heavy) hydrogen atoms and ${n_{h}}$ the number 
of the hydrogen atoms, with $\eta$ which might be bellow $10^{2}$.  
The  best measurements in the context of such baryons with the masses of a few hundred TeV/${\rm c}^2$  
which we were able to find were done $25$ years ago~\cite{superheavy}. The authors declare 
that their measurements manifest that such a ratio should be $\frac{n_{ah}}{n_{h}}< 10^{-14}$ for the mass interval 
between $10$ TeV/$c^{2}$ to $10^{4}$ TeV/$c^{2}$. Our evaluation presented above is very rough and more careful 
treating the problem might easily lead to lower values than required. On the other side we can not say 
how trustable is the value for the above ratio %(which confidence level it has) 
for the masses of a few hundreds 
TeV. Our evaluations are very approximate and if $\eta= 10^{2}$ we  conclude that the evaluation 
agrees with measurements.  
%%%%
%%%%
%(****TO BE CALCULATED AND CHECKED BY GREGOR, ALSO THE REFERENCE AND COMMENT ARE MISSING and will be added by him.****) 
%%%%
%%%%
  
% The de Broglie wavelength is $\approx 10 $ times the average distance among neighbour quarks.
%(The same would happen to all the lower 
%families' quarks and anti-quarks (going due to the Yukawa couplings to the first family members), 
%if there would be no   quark---anti-quark asymmetry, but mostly at the very end of this period, 
%when the temperature falls bellow $1$ MeV).  

%%POPRAVI CONCLUSION AND INTRODUCTION TER POVEJ OCENO, KOLIKO DOGODKOV BI IZMERILI CDMS PRI MASSI 100 TeV

%%END EVOLUTION

%
\section{Dynamics of a heavy family baryons in our galaxy}
\label{dynamics}

%

%%%
There are experiments~\cite{rita0708,cdms} which are trying to directly measure the dark matter clusters. Let us 
make a short introduction into these measurements, treating our fifth family clusters in particular.
The density of the dark 
matter $\rho_{dm}$ in the Milky way can be evaluated from the measured rotation velocity  
of  stars and gas in our galaxy, which appears to be approximately independent of the distance $r$ from the 
center of our galaxy. For our Sun this velocity 
is $v_S \approx (170 - 270)$ km/s. $\rho_{dm}$ is approximately spherically symmetric distributed 
and proportional to $\frac{1}{r^2}$.  Locally (at the position of our Sun) $\rho_{dm}$ 
is known within a factor of 10 to be 
$\rho_0 \approx 0.3 \,{\rm GeV} /(c^2 \,{\rm cm}^3)$, 
we put $\rho_{dm}= \rho_0\, \varepsilon_{\rho},$ 
with $\frac{1}{3} < \varepsilon_{\rho} < 3$. 
The local velocity distribution of the dark matter cluster $\vec{v}_{dm\, i}$, in the 
velocity  class 
$i$ of clusters, can only be estimated, 
results depend strongly on the model. Let us illustrate this dependence.  
%It is taken usually  as zero in the coordinate system fixed on 
%the center of our  galaxy. 
In a simple model that all the clusters at any radius $r$ from the center 
of our galaxy travel in all possible circles around the center so that the paths are 
spherically symmetrically distributed, the velocity of a cluster at the position of 
the Earth is equal to $v_{S}$, the velocity of our Sun in the absolute value,
but has all possible orientations perpendicular to the radius $r$ with  equal probability.
In the model %~\cite{gnBled07} 
that the clusters only oscillate through the center of the galaxy, 
the velocities of the dark matter clusters at the Earth position have values from 
zero to the escape velocity, each one weighted so that all the contributions give  
$ \rho_{dm} $. %Also the model  that clusters make all possible paths 
%from the oscillatory one to the circle, weighted so that they reproduce the $\rho_{dm}$,
%seems acceptable. 
Many other possibilities are presented in the references cited in~\cite{rita0708}. 

The velocity of the Earth around the center of the galaxy is equal to:  
$\vec{v}_{E}= \vec{v}_{S} + \vec{v}_{ES} $, with $v_{ES}= 30$ km/s and 
$\frac{\vec{v}_{S}\cdot \vec{v}_{ES}}{v_S v_{ES}}\approx \cos \theta \, \sin \omega t, \theta = 60^0$. 
Then the velocity with which the dark matter cluster of the $i$- th  velocity class  
hits the Earth is equal to:  
$\vec{v}_{dmE\,i}= \vec{v}_{dm\,i} - \vec{v}_{E}$. % where the index $i$ points out 
%that  the distribution in the velocity, which is very model dependent, is in 
%the class $i$. 
$\omega $ 
determines the rotation of our Earth around the Sun.

One finds for the flux %per unit time and unit surface 
of the  
%(any heavy with the small enough cross section) 
dark matter clusters hitting the Earth:    
$\Phi_{dm} = \sum_i \,\frac{\rho_{dm \,i}}{m_{c_5}}  \,
|\vec{v}_{dm \,i} - \vec{v}_{E}|  $ 
to be approximately  (as long as $\frac{v_{ES}}{|\vec{v}_{dm \,i}- \vec{v}_S|}$ is small%than $1$
) equal to  
\begin{eqnarray}
\label{flux}
\Phi_{dm}\approx \sum_i \,\frac{\rho_{dm \,i}}{m_{c_5}}  \,
\{|\vec{v}_{dm \,i} - \vec{v}_{S}| - \vec{v}_{ES} \cdot \frac{\vec{v}_{dm\, i}- \vec{v}_S}{
|\vec{v}_{dm \,i}- \vec{v}_S|} \}.
\end{eqnarray}
Further terms are neglected. %The flux is very much model dependent. 
We shall approximately take that
$\sum_i \, |\vec{v_{dm \,i}}- \vec{v_S}| \,\rho_{dm \,i} \approx \varepsilon_{v_{dmS}} 
\, \varepsilon_{\rho}\,  v_S\, \rho_0 $, %with $\rho_0 = 0.3 \, {\rm GeV}/(c^2\, cm^3), $ 
%while we estimate $\frac{1}{4} < \varepsilon_{\rho} < 4$,   
%$ \frac{1}{3} < \varepsilon_{v_{dmS}} < 3$ 
and correspondingly 
$ \sum_i \, \vec{v}_{ES}  \cdot \frac{\vec{v}_{dm \,i}- \vec{v}_S}{
|\vec{v}_{dm \,i}- \vec{v}_S|} \approx v_{ES} \varepsilon_{v_{dmS}}
\cos \theta \, \sin \omega t $, % with $v_{ES} = 30$ km/s  
%$\theta = 60^0$, 
(determining the annual modulations observed by DAMA~\cite{rita0708}). 
Here $\frac{1}{3} < \varepsilon_{v_{dmS}} < 
3$ and $\frac{1}{3} < \frac{\varepsilon_{v_{dmES}}}{\varepsilon_{v_{dmS}}} < 3$ are
estimated with respect to experimental and (our) theoretical evaluations.  

Let us evaluate the cross section for our heavy dark matter baryon to elastically
(the excited states of nuclei,  
which we shall treat, I and Ge, are at $\approx 50$ keV 
or higher and are very narrow, while the average recoil energy of Iodine is expected to be 
$30$ keV) 
scatter  on an ordinary nucleus with $A$ nucleons 
$\sigma_{A} = 
\frac{1}{\pi \hbar^2} <|M_{c_5 A}|>^2 \, m_{A}^2$. 
For our heavy dark matter cluster %with a small cross section $\sigma_{c_{5}}$  
is  $m_{A}  $  approximately the mass of the ordinary nucleus~\footnote{Let us illustrate 
what is happening when a very heavy ($10^4$ times or more heavier than the ordinary nucleon) cluster 
hits the nucleon. Having the "nuclear force" cross section of $10^{-8}$ ${\rm fm}^2$ or smaller, 
it "sees" with this cross section a particular quark, which starts to move. But since at this velocities 
the quark is tightly bound into a nucleon and nucleon into the nucleus, the hole nucleus is forced to move 
with the moving  quark.}. 
%%%%
%%%%KOMENTIRAJ, KAKO ZADENE TEZAK DELEC KVARK, NAPNE STRUNE IN PREMAKNE CELO JEDRO
%%%%
In the case of a 
coherent scattering (if recognizing that $\lambda= \frac{h}{p_A}$ is for a nucleus large enough 
to make scattering coherent when the mass of  the cluster is 
 $1$ TeV or more and its velocity 
$\approx v_{S}$), the cross section is  almost independent of the recoil 
velocity of the nucleus. 
For the case that the ''nuclear force'' as manifesting  in the cross section $\pi\, (r_{c_5})^2$ 
in Eq.(\ref{bohr}) 
brings the main contribution~\footnote{The very heavy colourless cluster of three quarks,  
hitting with the relative velocity $\approx 200 \,{\rm km}/{\rm s}$ the nucleus of the first 
family quarks, ''sees'' the (light) quark  $q_1$ of the 
nucleus through the cross section $\pi\, (r_{c_5})^2$.
But since the quark $q_{1}$ is at these velocities strongly bound to the proton and the 
proton to the nucleus,  the hole nucleus takes the momentum.} 
the cross section  is  proportional to $(3A)^2$ 
(due to the square of the matrix element) times $(A)^2$ (due to the mass of the nuclei 
$m_A\approx 3 A \,m_{q_1}$, with $m_{q_1}\, c^2 \approx \frac{1 {\rm GeV}}{3}$).  
When $m_{q_5}$ is  heavier than $10^4 \, {\rm TeV}/c^2$ (Table~\ref{TableI.}), 
the weak interaction dominates and $\sigma_{A}$ is proportional to $(A-Z)^2 \, A^2$, 
since to $Z^0$ boson exchange only neutron gives an appreciable contribution. 
Accordingly we have,  when the ''nuclear force'' dominates,
$\sigma_A \approx \sigma_{0} \, A^4 \, \varepsilon_{\sigma}$, with 
$\sigma_{0}\, \varepsilon_{\sigma}$, which is $\pi r_{c_5}^2 \, 
\varepsilon_{\sigma_{nucl}} $ and with 
 $\frac{1}{30} < \varepsilon_{\sigma_{nucl}} < 30$.  
$\varepsilon_{\sigma_{nucl}}$
takes into account the roughness 
with which we treat our  heavy baryon's properties and the scattering procedure.  
When the weak interaction dominates, $ \varepsilon_{\sigma}$ is smaller and we have $  
 \sigma_{0}\, \varepsilon_{\sigma}=(\frac{m_{n_1} G_F}{\sqrt{2 \pi}} 
\frac{A-Z}{A})^2 \,\varepsilon_{\sigma_{weak}}  $
($=( 10^{-6} \,\frac{A-Z}{ A} \, {\rm fm} )^2 \,\varepsilon_{\sigma_{weak}} $), 
$ \frac{1}{10}\, <\,  \varepsilon_{\sigma_{weak}} \,< 1$. The weak force is pretty accurately 
 evaluated, but the way how we are averaging is not.

\section{Direct measurements of the fifth family  baryons as dark matter constituents} 
\label{directmeasurements}

We are making very rough estimations of what the  
 DAMA~\cite{rita0708} and CDMS~\cite{cdms} experiments are measuring, provided that the 
 dark matter clusters are made out 
 of our (any) heavy family quarks as discussed above. 
 We are looking for limitations these two experiments might put on 
 properties of our heavy family members. 
 We discussed about our estimations and their relations to the measurements 
 with R. Bernabei~\cite{privatecommRBJF} and 
 J. Filippini~\cite{privatecommRBJF}. 
 Both pointed out (R.B. in particular) that the two experiments can hardly be compared, 
 and that our very approximate estimations may be right only within the orders of magnitude. 
 % detailed taking into 
 %account  the way how do the dark matter constituents scatter on the nuclei and with 
 %which velocity do they scatter (in ref.~\cite{rita0708} such studies were done),   
 %as well as how does a particular  experiment measure events,  is essential and that 
 %results depend  significantly on the details, so that too rough treating might change the 
 %results for orders of magnitude. 
 We are completely aware of how rough our estimation is, % and we 
 %do agree with  their comments, 
 yet we conclude that, since the number of measured events  is  proportional to 
 $(m_{c_5})^{-3}$ %the third power of the clusters' mass, 
 for masses $\approx 10^4$ TeV or smaller (while for 
 higher masses, when the weak interaction dominates, it is proportional to  
 $(m_{c_5})^{-1}$) that even such rough  estimations   
 may in the case of our heavy baryons say whether both experiments
 do at all measure our (any) heavy family clusters, if one experiment 
 clearly sees  the dark matter signals and the 
 other does not (yet?) and we accordingly estimate the mass of our cluster. 
 % Let us point out that the number of events an experiment  on the Earth might 
 %recognize as triggered by our heavy  dark matter cluster  is proportional to 
 %$1/m_{c_5}^3$, so that accordingly the  mass estimated from the measured events 
 % depends on the third route of the number of events. 
 
 Let $N_A$ be the number of nuclei of a type $A$ in the %measurement 
 apparatus  
 (of either DAMA~\cite{rita0708}, which has $4\cdot 10^{24}$ nuclei per kg of $I$, 
 with $A_I=127$,  
  and  $Na$, with $A_{Na}= 23$ (we shall neglect $Na$), 
 or of CDMS~\cite{cdms}, which has $8.3 \cdot 10^{24}$ of $Ge$ nuclei 
 per kg,  with $A_{Ge}\approx 73$). 
 At velocities  of a dark matter cluster  $v_{dmE}$ $\approx$ $200$ km/s  
 are the $3A$ scatterers strongly bound in the nucleus,    
 so that the whole nucleus with $A$ nucleons elastically scatters on a 
 heavy dark matter cluster.  
Then the number of events per second  ($R_A$) taking place 
in $N_A$ nuclei   is  due to the flux $\Phi_{dm}$ and the recognition that the cross section 
is at these energies almost independent 
of the velocity %(and depends accordingly only  on $A$ of the nucleus),  
equal to
\begin{eqnarray}
\label{ra}
R_A = \, N_A \,  \frac{\rho_{0}}{m_{c_5}} \;
\sigma(A) \, v_S \, \varepsilon_{v_{dmS}}\, \varepsilon_{\rho} \, ( 1 + 
\frac{\varepsilon_{v_{dmES}}}{\varepsilon_{v_{dmS}}} \, \frac{v_{ES}}{v_S}\, \cos \theta
\, \sin \omega t).
\end{eqnarray}
Let $\Delta R_A$ mean the amplitude of the annual modulation of $R_A$ 
\begin{eqnarray}
\label{anmod}
\Delta R_A &=& R_A(\omega t = \frac{\pi}{2}) - R_A(\omega t = 0) = N_A \, R_0 \, A^4\, 
\frac{\varepsilon_{v_{dmES}}}{\varepsilon_{v_{dmS}}}\, \frac{v_{ES}}{v_S}\, \cos \theta,
\end{eqnarray}
where $ R_0 = \sigma_{0} \, \frac{\rho_0}{m_{c_5}} \,  v_S\, \varepsilon$, 
$R_0$ is for the case that the ''nuclear force'' 
dominates $R_0 \approx  \pi\, (\frac{3\, \hbar\, c}{\alpha_c \, m_{q_5}\, c^2})^2\, 
\frac{\rho_0}{m_{q_5}} \, v_S\, \varepsilon$, with 
$\varepsilon = 
\varepsilon_{\rho} \, \varepsilon_{v_{dmES}} \varepsilon_{\sigma_{nucl}} $.  $R_0$ is therefore 
proportional to $m_{q_5}^{-3}$. 
We estimated  $10^{-4} < \varepsilon < 10$,   %(****ALI LAHKO TO NAPAKO ZMANJSAMO?****) 
which demonstrates both, the uncertainties in the knowledge about the dark matter dynamics 
in our galaxy and our approximate treating of the dark matter properties.  
(When for $m_{q_5} \, c^2 > 10^4$ TeV the weak interaction determines the cross section  
$R_0 $ is in this case proportional to $m_{q_5}^{-1}$.) 
We estimate that an experiment with $N_A$ scatterers  should  measure the amplitude
$R_A \varepsilon_{cut\, A}$, with $\varepsilon_{cut \, A}$ determining  the efficiency  of 
a particular experiment to detect a dark matter cluster collision. 
For small enough $\frac{\varepsilon_{v_{dmES}}}{\varepsilon_{v_{dmS}}}\, 
\frac{v_{ES}}{v_S}\, \cos \theta$ we have 
\begin{eqnarray}
R_A \, \varepsilon_{cut \, A}  \approx  N_{A}\, R_0\, A^4\, 
 \varepsilon_{cut\, A} = \Delta R_A \varepsilon_{cut\, A} \,
 \frac{\varepsilon_{v_{dmS}}}{\varepsilon_{v_{dmES}}} \, \frac{v_{S}}{v_{ES}\, \cos \theta}. 
\label{measure}
\end{eqnarray}
If DAMA~\cite{rita0708}   is measuring 
our  heavy  family baryons %with weak enough scattering cross section 
%(mostly)  
 scattering mostly on $I$ (we neglect the same number of $Na$,  with $A =23$),  
then the average $R_I$ is 
\begin{eqnarray}
\label{ridama}
R_{I} \varepsilon_{cut\, dama} \approx  \Delta R_{dama} % \; \varepsilon_{cut\, dama}\,
\frac{\varepsilon_{v_{dmS}}}{\varepsilon_{v_{dmES}}}\,
\frac{v_{S}  }{v_{ES}\, \cos 60^0 } ,
\end{eqnarray}
with $\Delta R_{dama}\approx 
\Delta R_{I}  \, \varepsilon_{cut\, dama}$, this is what we read from their papers~\cite{rita0708}.  
In this rough estimation 
most of unknowns about the dark matter properties, except the local velocity of our Sun,  
the cut off procedure ($\varepsilon_{cut\, dama}$) and 
$\frac{\varepsilon_{v_{dmS}}}{\varepsilon_{v_{dmES}}}$,
(estimated to be $\frac{1}{3} < \frac{\varepsilon_{v_{dmS}}}{\varepsilon_{v_{dmES}}} < 3$), 
 are hidden in $\Delta R_{dama}$. If we assume that the 
Sun's velocity is 
$v_{S}=100, 170, 220, 270$ km/s,  we find   $\frac{v_S}{v_{ES} \cos \theta}= 7,10,14,18, $ 
respectively. (The recoil energy of the nucleus $A=I$ changes correspondingly %in the average 
with the square of   $v_S $.)
DAMA/NaI, DAMA/LIBRA~\cite{rita0708} publishes %with $4. 10^{24}$ scatterers  per kg 
%$\varepsilon_{cut \,I}$ \, 
$ \Delta R_{dama}= 0.052  $ counts per day and per kg of NaI. 
Correspondingly  is $R_I \, \varepsilon_{cut\, dama}  = 
 0,052 \, \frac{\varepsilon_{v_{dmS}}}{\varepsilon_{v_{dmES}}}\, \frac{v_S}{v_{SE} \cos \theta} $ 
counts per day and per kg. 
CDMS should then in $121$ days with 1 kg of Ge ($A=73$) detect   
$R_{Ge}\, \varepsilon_{cut\, cdms}$
$\approx \frac{8.3}{4.0} \, 
 (\frac{73}{127})^4 \; \frac{\varepsilon_{cut\,cdms}}{\varepsilon_{cut \,dama}}\, 
 \frac{\varepsilon_{v_{dmS}}}{\varepsilon_{v_{dmES}}}\;
 \frac{v_S}{v_{SE} \cos \theta} \;  0.052 \cdot 
 121 \;$ events, 
which is for the above measured velocities equal to $(10,16,21,25)
\, \frac{\varepsilon_{cut\, cdms}}{\varepsilon_{cut\,dama}}\;
\frac{\varepsilon_{v_{dmS}}}{\varepsilon_{v_{dmES}}}$. CDMS~\cite{cdms} 
has found no event.

The approximations we made might cause that the expected  numbers 
$(10,16,21,25)$ multiplied by $\frac{\varepsilon_{cut\,Ge}}{\varepsilon_{cut\,I}}\;
\frac{\varepsilon_{v_{dmS}}}{\varepsilon_{v_{dmES}}}$  
are too high (or too low!!) for a factor let us say $4$ or $10$. 
%(But they also might be too low for the same fastor!)  
If in the near future  
CDMS (or some other experiment) 
will measure the above predicted events, then there might be  heavy 
family clusters which form the dark matter. In this case the DAMA experiment   
puts the limit on our heavy family masses (Eq.(\ref{measure})). 

Taking into account all the uncertainties   mentioned above, with the uncertainty with
the ''nuclear force'' cross section included (we evaluate these uncertainties  to be 
$10^{-4}  <\,\varepsilon^{"}\,< 3\cdot 10^3$), we can estimate the mass range of the fifth family quarks 
from the DAMA experiments:  $(m_{q_5}\, c^2)^3= \frac{1}{\Delta R_{dama}} 
N_I\,A^4\, \pi \,(\frac{3 \,\hbar c}{\alpha_c})^2 \,
\rho_0\, c^2\, v_{ES} \,\cos \theta\, \varepsilon^{"}= (0.3\, \cdot 10^7)^3 \, 
\varepsilon^{"} (\frac{0.1}{\alpha_c})^{2}
$ GeV. The lower mass limit, which follows from the DAMA experiment,  is accordingly  
$m_{q_5}\, c^2> 200$ TeV. 
Observing that 
for $m_{q_5} \, c^2> 10^4$ TeV 
the weak force starts to dominate, we estimate the upper limit $m_{q_5}\, c^2< 10^5$ TeV. 
Then
%In the case that the weak 
%interaction determines the $n_5$ cross section we find for the mass range    
$200 {\rm\; TeV} < m_{q_5} \, c^2 < 10^5$ TeV. 

Let us at the end evaluate the total number of our fifth family neutrons ($n_5$) which in $\delta t= 121$ days 
strike $1$ kg of Ge and which CDMS experiment could detect, that is  $R_{Ge} \delta t 
\varepsilon_{cut_{Ge}}= N_{Ge} \sigma_0 \frac{\rho_0}{m_{c_5}}\,v_{S}\, A^{4}_{Ge}\, \varepsilon  \varepsilon_{cut+{Ge}}$ 
(Eq.~\ref{measure}), 
with $N_{Ge} = 8.3 \cdot 10^{24}$/kg, with the cross section from Table~\ref{TableI.}, with $A_{Ge} = 73$ 
and $1$ kg of Ge, while  $10^{-5} < \varepsilon \varepsilon_{cut_{Ge}}< 5\cdot 10$. The  coefficient 
$\varepsilon \varepsilon_{cut_{Ge}}$ determines all the uncertainties: about the scattering amplitudes of the 
fifth family neutrons on the Ge nuclei (about the scattering amplitude of one 
$n_5$ on the first family quark, about the degree of coherence when scattering on the nuclei, about the 
local density of the dark matter, about the local velocity of the dark matter and about the efficiency of the 
experiment). Quite a part of these uncertainties were hidden in the number of events the DAMA/LIBRA 
experiments measure, when we compare both experiments. 
If we assume that the fifth family quark mass ($m_{q_5}$) is several hundreds TeV, as evaluated (as  
the upper bound (Eq.~\ref{massinterval})) when considering the cosmological history of our fifth family neutrons,
we get for the number of events the CDMS experiment should measure: 
$\varepsilon \varepsilon_{cut_{Ge}} \cdot 10^{4}$. If we take $\varepsilon \varepsilon_{cut_{Ge}}= 10^{-5}$,
the CDMS experiment should continue to measure 10 times as long as they did.

Let us see how many events CDMS should measure if the dark matter clusters would interact weakly 
with the Ge nuclei and  if the 
weak interaction would determine also their freezing out procedure, that is if any kind of WIMP would 
form the dark matter. One easily sees from the 
Boltzmann equations for the freezing out procedure for $q_5$ that since the weak massless boson exchange 
is approximately hundred times weaker than the 
one gluon exchange which determines the freeze out procedure of the fifth family quarks,  
the mass of such an object should be hundred times smaller, which means a few TeV. Taking into account the 
expression for the weak interaction of such an object with Ge nuclei, which leads to $10^{-2}$ smaller 
cross section for scattering of one such weakly interacting particle on one proton (see derivations in the previous 
section), we end up with the 
number of events which the CDMS experiment should measure: $\varepsilon \varepsilon_{cut_{Ge}} 5 \cdot 10^3$. 
Since the weak interaction with the matter is much better known that the (''fifth family nuclear force'') 
interaction of the colourless clusters of $q_5$ ($n_5$), the 
$\varepsilon $ is smaller. Let us say $\varepsilon$ is $5 \cdot10^{-4}$. Accordingly, even in the case of 
weakly interacting 
dark matter particles the CDMS should continue to measure to see some events.

\section{ Concluding remarks}
\label{conclusion}

We estimated in this paper the possibility that a new  stable  family, 
predicted by the approach unifying spin and 
charges~\cite{pn06,n92,gmdn07} to have the same charges and the same couplings 
to the corresponding gauge fields as the known  families,    
%having the matrix elements of the Yukawa couplings to the lower mass 
%families equal to zero, 
%
forms baryons which are the dark matter constituents. The approach (proposed by S.N.M.B.)
 is to our knowledge the only proposal  
in the literature so far which offers the mechanism for generating families, 
if we do not count those which in one or another way just assume more 
than three families. 
Not  being able so far to derive  from the approach precisely enough the fifth family masses and also not 
(yet) the baryon asymmetry,
we assume that the neutron is the lightest fifth family baryon and that there is no baryon---anti-baryon asymmetry.
We comment what changes if the asymmetry exists.
%%%%
%POPRAVITI MORDA
%%%%
%%%%
We evaluated under these assumptions the properties of the fifth family members  in the expanding universe, 
their clustering into the fifth family 
neutrons, the scattering of these neutrons on ordinary matter and find the limit on the properties of the stable fifth family quarks 
due to the 
cosmological observations and the direct experiments provided that these neutrons constitute the dark matter.

We use the simple hydrogen-like model to evaluate the 
properties of these heavy baryons and their interaction %of heavy baryons %among themselves and 
among themselves and with the ordinary  nuclei. We take into account that for masses of the order 
of $1$ TeV/$c^2$ or larger the one gluon exchange determines the force among the constituents of 
the fifth family baryons. Studying the interaction of these baryons with the ordinary matter we 
find out  that %in the latter case 
for massive enough fifth family quarks ($m_{q_5}> 10^4$ TeV) the weak interaction 
starts to dominate over the 
''nuclear interaction'' which the fifth family neutron manifests.  
The non relativistic 
fifth family baryons interact among themselves with the weak force only.

%
%We assume %further (with no justification yet) 
%that in the evolution of our universe $q_5$ and $\bar{q}_5$ were formed with no asymmetry. 
We study  
the freeze out procedure of the fifth family quarks and anti-quarks and the formation of  
baryons and anti-baryons up to the temperature  $ k_b T= 1$ GeV,  when the colour phase transition 
starts which to our estimations depletes almost all the fifth family quarks and anti-quarks while the colourless
% (neutral with respect to the colour and electromagnetic charge) 
fifth family neutrons with very small scattering cross section decouples long before (at $ k_b T= 100$ GeV). 

The cosmological evolution 
suggests for the mass limits the range $10$ TeV $< m_{q_5} \, c^2 < {\rm a \, few} \cdot 10^2$ TeV 
and for the  scattering cross sections 
$ 10^{-8}\, {\rm fm}^2\, < \sigma_{c_5}\, <   10^{-6} \,{\rm fm}^2  $. 
The measured density of  the  dark matter 
does not put much limitation on the properties of heavy enough clusters.

The DAMA experiments~\cite{rita0708} limit (provided that they measure 
our heavy fifth family clusters) the quark mass 
to:  $ 200 \,{\rm TeV} < m_{q_{5}}c^2 < 10^5\, {\rm TeV}$.   
%In the case that the weak interaction determines the $n_5$ cross section we find     
%$10\; {\rm TeV} < m_{q_5} \, c^2< 10^5$ TeV.    
The estimated cross section for the dark matter cluster to 
(elastically, coherently and nonrelativisically) scatter on the (first family) nucleus is in this case 
determined on the lower mass limit by the ''fifth family nuclear force'' of the fifth family clusters %and is equal to 
($ (3\cdot 10^{-5}\,A^2\, {\rm fm} )^2$) % while 
and on the higher mass limit by the weak force %determines %the cross 
%section, which is equal to  
($ ( A (A-Z)\, 10^{-6} \, {\rm fm} )^2 $). 
Accordingly we conclude that if the DAMA experiments are measuring our fifth family neutrons,  
the mass of the fifth family quarks is a few hundred  TeV $/c^2$. 

Taking into account all the uncertainties in connection with the dark matter clusters (the local density of the 
dark matter and its local velocity) including the scattering cross sections  of our fifth family neutrons on the 
ordinary nuclei as well as the experimental errors, we do expect that CDMS will in a few years 
measure our fifth family baryons.

Let us point out that  the stable fifth family neutrons are not the WIMPS, which would interact with the weak force 
only: the 
cosmological behaviour (the freezing out procedure) of these clusters are dictated by the colour force, while their 
interaction with the ordinary matter is determined by the "fifth family nuclear force" if 
they have masses smaller than $10^4$ TeV/$c^2$.

In the ref.~\cite{mbb}~\footnote{ The referee of PRL suggested 
that we should comment on the paper~\cite{mbb}.} the authors 
study the limits on a scattering cross section of 
a heavy dark matter cluster of particles and anti-particles (both of approximately the same amount) 
with the ordinary matter, estimating the energy 
flux produced by the annihilation of such pairs of clusters. %They assume (approximately) the same number of 
%particles and antiparticles in the dark matter. 
They treat the conditions under which 
would  %in a stationary case  
the heat flow  following  
from the annihilation of dark matter  particles and anti-particles in the Earth core start to be noticeable. 
Using their limits we conclude that our fifth family baryons of the mass of a few hundreds TeV/${c^2} $ 
have  for a factor more than $100$ too small scattering amplitude with the ordinary matter to cause a measurable 
heat flux on the Earth's surface. 
On the other hand could the measurements~\cite{superheavy} tell whether the fifth family members do deplete 
at the colour phase transition of our universe enough to be in agreement with them. Our very rough estimation 
show that the fifth family  members are on the allowed limit, but they are too rough to be taken as 
a real limit.

%Let us add that our Earth would for $m_{q_5}\,c^2 \approx 1 {\rm TeV}$ or lower contain a mass part 
%$10^{-9}$ or lower of the dark matter clusters, and that  
%the mean time between two collisions among the dark matter clusters in our galaxy would be 
%from $10^{22}$ years on. 

Our estimations predict that, if the DAMA experiments %~\cite{rita0708} 
observe the events due to our (any) 
heavy family members, (or any heavy enough family clusters with 
small enough cross section),  
the CDMS experiments~\cite{cdms} will in the near future observe  
a few events as well. %(limiting now the heavy family mass
%to be $8. 10^3$ TeV or higher).  
%(in particular if clusters scatter elastically and coherently, which 
%is happening if their velocities are low enough).  
%The CDMS experiment limits up to now the heavy family mass
%to be $8. 10^3$ TeV or higher. 
%
If CDMS will not confirm the heavy family events, then we must conclude, 
trusting the DAMA experiments, that either our 
fifth family clusters have much higher cross section due to the possibility that $u_5$ is lighter than 
$d_5$  so that their velocity slows down when 
scattering on nuclei of the earth above the measuring apparatus 
bellow the threshold of the CDMS experiment (and that there must be in this case the fifth family 
quarks---anti-quarks asymmetry)~\cite{maxim}) 
while the DAMA experiment still observes them, 
%  
%other possibilities ($p_5, \Delta_{5}^{++}, \Delta_{5}^{-}, \bar{\Delta}_{5}^{++} $) might fit 
%the data, % ~\cite{maxim} 
or the fifth family clusters (any heavy stable family clusters) are not what forms the dark matter.

Let us comment again the question whether it is at all possible (due to electroweak experimental
data) that there exist more than three up to now observed families, that is, whether the approach 
unifying spin and charges  by predicting the fourth and the stable fifth  
family (with neutrinos included) contradict the observations. In the 
ref.~\cite{mdnbled06} the properties 
of all the members of the fourth family were studied (for  one particular choice of breaking the starting 
symmetry). The predicted fourth family neutrino mass is at around $100$ GeV/$c^2$ or higher, therefore it 
does not due to the detailed analyses of the electroweak data done by the Russian group~\cite{okun} 
contradict any experimental data. %(%****DODAJ REFERENCO STRAN DATA PRL****).  
The stable fifth family neutrino has due to our calculations %~\cite{mdnbled06} 
considerably higher mass. Accordingly none of these two neutrinos contradict  
the electroweak data. They also do not 
contradict the nucleosynthesis, since to the nucleosynthesis only the neutrinos with masses 
bellow the electron mass contribute. 
The fact that the fifth family baryons might form the dark matter does not contradict  
 the measured (first family) baryon number and its ratio to the photon 
 energy density as well, as long as the fifth family quarks are 
 heavy enough ($>$1 TeV). All the measurements, which connect the baryon and the photon 
 energy density, relate to the moment(s) in the history of 
 the universe, when the baryons (of the first family) where formed ($m_1 c^2 \approx 
  k_b T = 1$ GeV and lower)  and the electrons and nuclei were  forming  atoms ($k_b \,T 
 \approx 1$ eV). The chargeless (with respect to the colour and electromagnetic 
 charges, not with respect to the weak charge) clusters of the fifth family were 
 formed long before (at $ k_b T\approx E_{c_5}$ (Table~\ref{TableI.})). They  manifest 
 after decoupling from the plasma (with their small number density and  small cross 
 section) (almost) only their gravitational  interaction.

%If   future results from CDMS and 
%DAMA and other experiments will confirm our heavy family clusters with no light family quarks contributing,   
%then we shall soon know, what is the origin of the dark matter.  
%If a possible answer is a complicated mechanism %(like in ~\cite{maxim}),  
%then, since there might be  many possible complicated scenarios for the 
%dark matter origin, it might be very difficult to make the right choice among them, and 
%we shall not find out very soon what is the dark matter constituted out of.

  Let the reader recognize that the fifth family baryons are not the objects---WIMPS---which 
  would interact with only the weak interaction, 
 since their decoupling from the rest of the 
 plasma in the expanding universe is determined by the colour force and  
 their interaction with the ordinary matter is determined with the  
 fifth family "nuclear force" (the force among the fifth family nucleons, 
 manifesting much smaller cross section than does the ordinary 
 "nuclear force") as long as their mass is not higher than $10^{4} $ TeV, when the weak interaction starts to 
 dominate as commented in %the last paragraph of 
 section~\ref{dynamics}.

Let us conclude this paper with the recognition:   
 If the approach unifying spin and charges is the right way beyond the 
 standard model of the electroweak and colour interaction,  
 then more than three 
 families of quarks and leptons do exist, and the stable 
 (with respect to the age of the universe) fifth family of quarks and leptons 
 is the candidate to form the dark matter. The assumptions we made (i. The fifth 
 family neutron is the lightest fifth family baryon, ii. There is no fifth family baryon asymmetry),
 could be derived from the approach unifying spins and charges and we are working on these problems. 
 The fifth family baryon anti-baryon asymmetry does not very much  change the conclusions of 
 this paper as long as the fifth family quarks's mass is  a few hundreds TeV or higher.

 \section{ Appendix I. Three fifth family quarks' bound states}
 \label{betterhf}

 We look for the ground  state solution of the Hamilton equation  $H\,
 |\psi\rangle= E_{c_5}\,|\psi\rangle $ for a cluster of three heavy quarks with 
 \begin{eqnarray}
  H=\sum_{i=1}^3 \,\frac{p_{i}^2}{2 \,m_{q_5}} 
  -\frac{2}{3}\,  \, \sum_{i<j=1}^3
   \frac{\hbar c \; \alpha_c}{|\vec{x}_i-\vec{x}_j|},   
 \end{eqnarray}
 in the center of mass motion
 \begin{eqnarray}
  \vec{x}=\vec{x}_2 - \vec{x}_1,\quad  \vec{y}=\vec{x}_3 - \frac{\vec{x}_1+\vec{x}_2}{2},\quad
  \vec{R}=\frac{\vec{x}_1+\vec{x}_2+\vec{x}_3}{3}, 
 \end{eqnarray}
 assuming the anti-symmetric colour part  ($|\psi\rangle_{c,\, \cal{A} }$), 
 symmetric spin and weak charge part  ($|\psi\rangle_{w  \, {\rm spin},\, \cal{S}  }$)
 and symmetric space part ($|\psi\rangle_{{\rm space}, \, \cal{S}}$). 
 For the space part we take the hydrogen-like wave functions 
 $ \psi_a(\vec{x})=$$\frac{1}{\sqrt{\pi a^3}} \; e^{-|\vec{x}|/a}$ and 
  $\psi_b(\vec{y})=$$\frac{1}{\sqrt{\pi b^3}} \; e^{-|\vec{y}|/b}$, allowing $a$ and $b$ to 
  adapt variationally.
  Accordingly $\langle\, \vec{x}_1, \vec{x}_2, \vec{x}_3|\psi\rangle_{{\rm space}\, \cal{S}}=
  \mathcal{N}
 \left( \psi_a(\vec{x}) \psi_{b}(\vec{y}) + \textrm{symmetric permutations} \right)$. It follows 
 $\langle\, \vec{x}_1, \vec{x}_2, \vec{x}_3|\psi\rangle_{{\rm space}\, \cal{S}}=
 \mathcal{N} \,  \left(  2 \psi_a(\vec{x}) \psi_{b}(\vec{y})   +
  2 \psi_a(\vec{y}-\frac{\vec{x}}{2}) \psi_{b}(\frac{\vec{y}}{2}+\frac{3 \vec{x}}{4}))  +
 2 \psi_a(\vec{y}+\frac{\vec{x}}{2}) \psi_{b}(\frac{\vec{y}}{2}-\frac{3 \vec{x}}{4}) \right) $.
 
 The Hamiltonian in the center of mass motion reads 
 $H=\frac{p_x^2}{2 (\frac{m_{q_5}}{2})}+\frac{p_y^2}{2 (\frac{2m_{q_5}}{3})}+\frac{p_R^2}{2 
 \cdot 3 m_{q_5}}
 -\frac{2}{3} \hbar c \; \alpha_c \left(\frac{1}{x}+\frac{1}{|\vec{y}+\frac{\vec{x}}{2}|}+
 \frac{1}{|\vec{y}-\frac{\vec{x}}{2}|} \right).
 $
 Varying the expectation value of the Hamiltonian with respect to $a$ and $b$ 
 it follows: $\frac{a}{b}=1.03, \, \frac{a\, \alpha_c\, m_{q_5}\, c^2}{\hbar c} = 1.6$. 
 
 Accordingly we get for the binding energy  $ E_{c_5}=0.66\; m_{q_5}\, c^2 \alpha_{c}^2$ and for the size 
 of the cluster $\sqrt{\langle |\vec{x}_2-\vec{x}_1|^2 \rangle} = 2.5\, \frac{\hbar c}{\alpha_c m_{q_5 \, c^2}}
 $.

 To estimate  the  mass difference between $u_5$ and $d_5$
 for which $u_5 d_5 d_5$ is stable we treat the electromagnetic ($\alpha_{elm}$) and weak ($\alpha_w $) 
 interaction as a small correction 
 to the above calculated binding energy: $H'=  \alpha_{elm\,w}  \; \hbar c \,
 \left(\frac{1}{x}+\frac{1}{|\vec{y}+\frac{\vec{x}}{2}|}+
 \frac{1}{|\vec{y}-\frac{\vec{x}}{2}|} \right)$. $\alpha_{elm\,w} $ stays for electromagnetic and 
 weak coupling constants.
 For $m_{q_5}= 200$ TeV we take $\alpha_{elm\,w} =\frac{1}{100}$, then 
 $|m_{u_5}- m_{d_5}|< \frac{1}{3}\,  E_{c_5} \frac{(\frac{3}{2} \alpha_{elm\,w})^2}{\alpha_c^2}
 = 0.5\,\cdot 10^{-4} \; m_{q_5}\,  c^2 $.

 %
 %\appendix{Boltzmann equations for fifth family quarks and clusters}
%\label{beq}
%

%
\section*{Acknowledgments} 

The authors  would like to thank  all the participants 
of the   workshops entitled 
"What comes beyond the Standard models", 
taking place  at Bled annually (usually) in  July, starting in 1998, and in particular  
H. B. Nielsen, since all the open problems were there very openly discussed.

%pn06,n92,n93,n07bled,hn02hn03

\end{document}